\documentclass[compsoc, conference, a4paper, 10pt, times]{IEEEtran}
\PassOptionsToPackage{x11names,svgnames}{xcolor}

\usepackage[utf8]{inputenc}
\usepackage[english]{babel}
\usepackage[T1]{fontenc}
\usepackage{lmodern}

\usepackage{hyphenat}
\usepackage{hyperref}
\usepackage{diagbox}
\usepackage[finalizecache=false,frozencache=true]{minted}

\usepackage{layouts}

\hypersetup{
	backref=page,
	colorlinks,
	linkcolor={black},
	citecolor={red!70!black},
	urlcolor={blue!70!black}
}

\usepackage{multirow}
\usepackage{tabularx}

\usepackage[backend=biber,
	sortcites=true,
	maxbibnames=10,
	backref=true,
]{biblatex}
\AtBeginBibliography{\footnotesize}

\usepackage{times}

\usepackage[]{graphicx}
\usepackage{xspace}
\usepackage{amssymb}
\usepackage{tabulary}
\usepackage{pifont}
\PassOptionsToPackage{hyphens}{url}
\usepackage{float}
\usepackage{adjustbox}
\usepackage{booktabs}
\usepackage{wasysym}
\usepackage{xstring}
\usepackage{soul}
\usepackage[olditem,oldenum]{paralist}
\usepackage{cleveref}
\usepackage{csquotes}
\usepackage[binary-units=true,group-separator={,}]{siunitx}
\usepackage{relsize}
\usepackage{caption}
\usepackage{subcaption}

\AtBeginDocument{
  \DeclareSIUnit\bit{{bit}}%
  \DeclareSIUnit\byte{{byte}}%
  \DeclareSIUnit\day{{days}}%
}

\usepackage{todonotes}

\definecolor{lstgreen}{rgb}{0,0.7,0}
\usepackage{listings}
\lstdefinelanguage{solidity}{%
  morekeywords={new, true, false, function, return,
  switch, var, if, in, while, do, else, case, break,
  contract, mapping, struct, library,
  revert, assert, require, throw, sha3, keccak256,
  sstore, sload, delete},
  morekeywords=[2]{export, public, private, payable, view, returns},
  morekeywords=[3]{bool, uint, int, address, uint256, int256, int8, uint8,
    uint128, int128, bytes, bytes32, bytes4},
  morekeywords=[4]{msg, this, tx},
  sensitive=false,
  comment=[l]{//},
  morecomment=[s]{/*}{*/},
  morestring=[b]',
  morestring=[b]"
}
\lstdefinelanguage{evm}{%
  morekeywords={%
    push1, push2, push3, push32, not, and, or, jump, jumpi, swap1, swap2, push4, jumpdest, add, sub, mul, div, pop, invalid, dup1, dup2, dup3, dup8, stop, lt, iszero, revert, sha3, sload, sstore,
  },
  sensitive=false,
  comment=[l]{//},
  morecomment=[s]{/*}{*/},
}

\definecolor{lstgreen}{rgb}{0,0.6,0}
\usepackage{listings}
\lstdefinestyle{plain}{%
  numbers=none,
  frame=none,
  xleftmargin=1pt,
  xrightmargin=1pt,
}
\Crefname{lstlisting}{Listing}{Listings}

\usepackage{tikz}
\usetikzlibrary{shapes,arrows,positioning,shapes.multipart}

\usepackage{xspace}
\newcommand{\circled}[1]{\ding{\the\numexpr#1+191}\xspace}

\setminted{
	frame=lines,
	framesep=2mm,
	fontsize=\scriptsize,
	stripnl=true,
	linenos,
	breaklines=true,
	xleftmargin=1.2em,
	numbersep=8pt,
}

\newcommand{\toolname}{EF{\lightning}CF\xspace}
\newcommand{\toolnameslash}{EF/CF\xspace}
\newcommand{\toolnamelong}{Extremely Fast {\lightning} Contract Fuzzer\xspace}

\renewcommand{\paragraph}[1]{\medskip\noindent\textbf{#1}\xspace}

\newcommand{\citewauthor}[1]{\citeauthor*{#1}~\cite{#1}}

\newcommand{\p}[1]{\SI{#1}{\percent}}
\newcommand{\s}[1]{\SI{#1}{\second}}

\DeclareSIUnit\wei{Wei}
\DeclareSIUnit\ether{Ether}
\DeclareSIUnit\gas{gas}
\DeclareSIUnit\loc{loc}
\DeclareSIUnit\exec{exec}
\DeclareSIUnit\tx{tx}

\newcommand{\Cpp}{\mbox{C\Rplus\Rplus}\xspace}
\renewcommand{\Cpp}{C++\xspace}

\newcommand{\token}[1]{\todo[size=\small,color=gray,inline]{Responsible: #1}}

\makeatletter%
\if@todonotes@disabled%
\else%
\fi
\makeatother

\IEEEoverridecommandlockouts

\begin{document}
\title{{\toolname}: High Performance Smart Contract Fuzzing for Exploit Generation}
\author{
	\IEEEauthorblockN{%
    Michael Rodler$^1$\thanks{$^1$ Work conducted at University of Duisburg-Essen before joining Amazon.}, David Paaßen$^2$, Wenting Li$^3$, Lukas Bernhard$^4$,\\
    Thorsten Holz$^5$, Ghassan Karame$^4$, Lucas Davi$^2$%
  }
  \IEEEauthorblockA{%
    \rm $^1$ Amazon Web Services, 
    \rm $^2$ University of Duisburg-Essen, 
    \rm $^3$ NEC Laboratories Europe, \\
    \rm $^4$ Ruhr-University Bochum, 
    \rm $^5$ CISPA Helmholtz Center for Information Security \\
    \textit{m@mrodler.eu}, 
    \textit{\{david.paassen, lucas.davi\}@uni-due.de},
    \textit{wenting.li@neclab.eu} \\
    \textit{\{lukas.bernhard, ghassan.karame\}@rub.de},
    \textit{holz@cispa.de}%
  }
}

\newcommand{\anonLink}{\url{https://www.dropbox.com/sh/and242i04ucrfh0/AABwzDDWMGAz05kAu69ec2f4a}}
\newcommand{\releaseLink}{\url{https://github.com/uni-due-syssec/efcf-framework/}}

\maketitle

\begin{abstract}%
Smart contracts are increasingly being used to manage large numbers of high-value cryptocurrency accounts.
There is a strong demand for automated, efficient, and comprehensive methods to detect security vulnerabilities in a given contract.
While the literature features a plethora of analysis methods for smart contracts, the existing proposals do not address the increasing complexity of contracts.
Existing analysis tools suffer from false alarms and missed bugs in today's smart contracts that are increasingly defined by complexity and interdependencies.
To scale accurate analysis to modern smart contracts, we introduce \toolname, a high-performance fuzzer for Ethereum smart contracts.
In contrast to previous work, \toolname efficiently and accurately models complex smart contract interactions, such as reentrancy and cross-contract interactions, at a very high fuzzing throughput rate.
To achieve this, \toolname trans\-piles smart contract bytecode into native C++ code, thereby enabling the reuse of existing, optimized fuzzing toolchains.
Furthermore, \toolname increases fuzzing efficiency by employing a structure-aware mutation engine for smart contract transaction sequences and using a contract's ABI to generate valid transaction inputs.
In a comprehensive evaluation, we show that \toolname scales better---without compromising accuracy---to complex contracts compared to state-of-the-art approaches, including other fuzzers, symbolic/concolic execution, and hybrid approaches.
Moreover, we show that \toolname can automatically generate transaction sequences that exploit reentrancy bugs to steal Ether.

\end{abstract}

\section{Introduction}%
\label{sec:intro}

Ethereum is the most prominent blockchain platform that supports \emph{smart contracts}: Programs that are stored and run as part of the blockchain protocol.
Smart contracts are the backbone of the emerging decentralized finance (DeFi) industry.
Due to its popularity, the security of Ethereum and its smart contracts layer has received considerable attention from the research community and industry.
Since the first high profile attack against ``the DAO'' contract~\cite{dao-attack}, the community has observed a continuous stream of attacks against smart contracts~\cite{ZhouYXCY020eeg,Torres21eyeofhorus}.

Analyzing smart contract code is challenging due to its stateful nature and the large number of potential bug classes.
Prior work on identifying vulnerabilities in smart contracts relied on various techniques, such as symbolic execution~\cite{Luu2016oyente,KruppR18teether,NikolicKSSH18maian}, model checking~\cite{KalraGDS18zeus,FrankAH20ethbmc}, and static analysis~\cite{Tsankov2018securify,SchneidewindGSM20ethor}.
While these methods are promising, many tools primarily identify potential bugs with heuristics and do not give proof of exploitability, or they suffer from scalability issues (e.g., due to state explosion).
Fuzz testing~\cite{NguyenP0L020sfuzz,HeBATV19ilf,Jiang18contractfuzzer,GriecoSCFG20echidna,Wustholz2020harvey,Choi2021smartian} has emerged as a promising method for detecting smart contract bugs.
Fuzzing-based approaches are able to generate the specific inputs that allow a developer to trigger and analyze the identified problem within a debugging environment.
However, existing fuzzing approaches suffer from various drawbacks, most notably:
\begin{inparaenum}[(1)]
	\item They do not scale to complex contracts that are used on the Ethereum blockchain today.
	We identify a clear trend that smart contracts are becoming more complex over time and therefore require more complex transaction sequences for comprehensive analysis (see \Cref{sec:problem}).
	\item They do not accurately model the complex interactions that are possible in Ethereum, such as reentrancy and cross-contract interactions.
	Especially reentrancy attacks have emerged as one of the most critical security issues, as demonstrated by multiple high-profile incidents over the last years~\cite{dao-attack, Torres21eyeofhorus, creamfinance, revestfinance}.
	However, existing analysis tools over-approximate reentrancy attacks, leading to inaccurate analysis results (see \Cref{sec:eval:bugdetection}).
\end{inparaenum}

In this paper, we tackle all of these challenges by introducing \emph{\toolnamelong} (\toolname), an optimized fuzzing framework and exploit generator.
Comprehensively analyzing smart contracts is challenging because their behavior depends on their \emph{internal state}, which changes depending on the order and parameters of called functions.
Even for moderately complex contracts, it is not feasible for an analysis tool to simply execute all permutations of available functions.
Using high-throughput fuzzing guided by code coverage, \toolname efficiently searches the space of possible transaction sequences to identify those that expose a vulnerability.
To increase the fuzzing throughput, we propose to speed up Ethereum smart contracts by translating the Ethereum virtual machine (EVM) bytecode to equivalent \Cpp code and using a high-performance coverage-guided fuzzer.

\toolname utilizes a simple---yet powerful---bug oracle: Ether gains.
This allows \toolname to effectively act as an exploit generator for Ether stealing attacks.
However, Ether gains do not cover all current smart contract security issues.
In order to extend \toolname to cover smart contract specific bugs, \toolname allows developers to define custom bug oracles in their Solidity code.

Besides these bug classes, reentrancy issues remain one of the most challenging and critical vulnerabilities for Ethereum smart contracts~\cite{Torres21eyeofhorus,Cecchetti2021serif}.
Many existing analysis tools attempt to detect potential reentrancy using over-approximate analyses~\cite{Tsankov2018securify,FeistGG19slither,mythril,sereum,Torres2021confuzzius,Choi2021smartian}.
In contrast, \toolname accurately detects reentrancy vulnerabilities.
Instead of relying on heuristics, \toolname simulates the behavior of multiple attacker-controlled smart contracts.
Each test case generated by \toolname not only specifies the transactions, but also the behavior of the simulated attacker contracts.
This allows the coverage-guided fuzzing process to explore the behavior of the attacker-controlled contracts by attempting to execute reentrant calls and changing returned data of callbacks.
\toolname does not need an explicit bug oracle for reentrancy vulnerabilities and instead uses the Ether gains bug oracle.
This allows \toolname to automatically synthesize Solidity attack contracts that reproduce the reentrancy attack on a live blockchain.
Moreover, \toolname identifies compositional security issues~\cite{Cecchetti2021serif} that arise only when \emph{multiple} contracts are combined.
Before deploying a certain combination of contracts, \toolname can fuzz the combination to detect compositional security issues given a certain set of contracts.
We have performed a comprehensive evaluation of our fuzzing approach.
To this end, we instantiated \toolname with the AFL++ fuzzer~\cite{Fioraldi2020aflpp}.
We compare \toolname to current state-of-the-art analysis tools, such as Manticore~\cite{MossbergMHGGFBD19manticore}, Echidna~\cite{GriecoSCFG20echidna}, MAIAN~\cite{NikolicKSSH18maian}, teEther~\cite{KruppR18teether}, VeriSmart/SmartTest~\cite{So2021verismart,So2021verismartsmartest}, and Smartian~\cite{Choi2021smartian} with respect to their \emph{time-to-bug}.
Overall, we spent more than \num{1300} CPU days to evaluate and compare existing analysis tools and \toolname.
We find that \toolname is the only analysis tool capable of successfully analyzing all contracts in our benchmark and scales even to contracts requiring complex and long transaction sequences to trigger a bug.
In addition, we compare the performance of \toolname with the hybrid fuzzers ILF~\cite{HeBATV19ilf} and ConFuzzius~\cite{Torres2021confuzzius} with respect to code coverage.
Our experiments show that \toolname outperforms existing approaches when analyzing complex smart contracts according to several code complexity metrics.
We show that \toolname identifies \SI{99.9}{\percent} of the access control bugs that EthBMC~\cite{FrankAH20ethbmc} identified and that \toolname finds vulnerabilities in contracts that EthBMC could not analyze.
In addition, we compare \toolname with the symbolic analyzer Sailfish~\cite{bose2021sailfish} for reentrancy bugs, where we show that \toolname accurately detects those reentrancy issues that are exploitable.
Demonstrating the practicality of \toolname, we find that only \num{5} out of the \num{26} verified reentrancy bugs of Sailfish are actually prone to reentrancy attacks that allow the attacker to steal Ether.
Finally, we show that \toolname is able to generate concrete transaction sequences for the compositional reentrancy bugs in the contracts used in the evaluation of the Serif~\cite{Cecchetti2021serif} static analyzer tool.
Furthermore, \toolname is the first analysis tool that is capable of generating an exploit for the compositional reentrancy attack against the \emph{Uniswap}/\emph{IMBTC} contracts.

\paragraph{Contributions} In summary, our main contributions are:%
\begin{compactitem}
	\item We show how to leverage conventional software fuzzers to efficiently test smart contracts with coverage-guided fuzzing, allowing the fuzzer to scale to large and complex contracts.
	To achieve this, we propose a transpilation approach to accelerate the fuzzing of bytecode programs, such as the EVM.
	Our approach removes the interpreter by directly translating bytecode into native code, using \Cpp as an intermediate language (\Cref{sec:design}).
	We augment a conventional base fuzzer, such as \emph{AFL++}, with a custom mutator to implement smart contract specific mutation operations.
	\item We efficiently model complex smart contract interactions during fuzzing, including multiple attacker-controlled smart contracts, reentrant calls, and cross-contract interactions in a fuzzer.
	Using this approach, \toolname can automatically generate sophisticated reentrancy exploits (\Cref{sec:implementation}).
	\item We thoroughly evaluate the performance of \toolname against a large number of state-of-the-art analysis tools (see \Cref{sec:eval}):
    \begin{inparaenum}[(a)]
      \item \toolname scales best to longer transaction sequences when compared using time-to-bug.
      \item \toolname achieves significantly better code coverage than prior fuzzers on existing real-world smart contracts, especially when considering various code complexity metrics.
      \item \toolname is capable of identifying access control and reentrancy bugs with high accuracy and fewer false alarms than current analysis tools.
    \end{inparaenum}
\end{compactitem}
To foster research on the security of smart contracts, we release \toolname along with all benchmarks and experiments at \releaseLink.

\section{Problem Statement}%
\label{sec:problem}

\paragraph{Ethereum Smart Contracts}
In Ethereum, each participant is identified by an address derived from cryptographic key material.
Each Ethereum account is defined by its address and an associated balance of Ether, Ethereum's native cryptocurrency.
Smart contract accounts are associated with a code and program state (called \emph{storage}).
An externally-owned account broadcasts a transaction to the Ethereum network to interact with a smart contract.
Transactions are used to either transfer Ether to trigger the execution of a smart contract or both.
A transaction consists of several fields, most importantly the destination address, the Ether value, and the input.
If the destination address is a smart contract, Ethereum nodes execute the smart contract with the provided input, and once a new Ethereum block is generated, the state updates of the transaction are included in the block and committed to the blockchain.

Smart contracts are written in specialized programming languages (e.g., Solidity) and compiled into Ethereum Virtual Machine (EVM) bytecode.
The EVM bytecode is a dedicated bytecode format optimized for small size, simplicity, and deterministic execution.
Most production-grade EVM implementations use a bytecode interpreter to execute a smart contract.
Ethereum smart contracts only have a single entry point. 
The executed high-level function is selected based on an identifier that is provided as part of the input.
The function parameters are encoded according to the \emph{Ethereum ABI} definition, which is a de-facto standard to encode parameters to function calls in the transaction input.

\paragraph{Challenges of Smart Contract Fuzzing}
When fuzzing smart contracts, the goal is to identify a sequence of transactions that exposes a software fault.
The final transaction of the sequence triggers a software fault, while the preceding transactions set up the state of the contract such that the fault can be triggered.
In this sense, testing Ethereum smart contracts is a variant of testing stateful software, a highly challenging problem~\cite{Claessen2000quickcheck,Kaksonen2001protos,Aitel2002spike,Xie2005symstra,Thummalapenta2011nz}.
In contrast to static analysis methods, generating complete transaction sequences by means of fuzzing has the advantage that it features a very low rate of false alarms, and the result is easy to analyze: A developer or security analyst can simply replay and debug the transaction sequence to determine the root cause and assess whether the bug can be triggered in practice.
However, determining such a transaction sequence---or function call sequence---is challenging since the search space is extremely large.
There are two dimensions that must be explored in parallel to reach high code coverage:
\begin{inparaenum}[(1)]
	\item the input to individual transactions and
	\item the ordering of the transactions.
\end{inparaenum}
To efficiently cover this large search space, we can exploit the fact that many of the possible transaction sequences are redundant since they only exercise the same error-handling paths over and over again.
\emph{Coverage-guided fuzzing} can efficiently search the input space of a given contract for inputs that trigger distinct code coverage and was popularized by the success of the American Fuzzy Lop (AFL) fuzzer~\cite{aflhomepage} and many follow-up works~\cite{Manes2021fuzzingsurvey,Fioraldi2020aflpp}.
In this context, test case throughput emerges as an important design aspect for an effective fuzzer.
Intuitively, the greater the number of test cases generated and executed, the greater the likelihood that a fault will be triggered within a given time budget during fuzzing, an inherently probabilistic process.
Most fuzz testing approaches for Ethereum smart contracts develop new fuzzers from scratch~\cite{HeBATV19ilf,GriecoSCFG20echidna,Torres2021confuzzius,NguyenP0L020sfuzz}, which are often not optimized for high throughput.
For example, ILF performs at a rate of 148 transactions per second~\cite{HeBATV19ilf}, while native code fuzzing with far more complex code regularly achieves \num{10000} or more test case executions per second~\cite{Xu2017fuzzprimitives}.
This low throughput effectively hampers a fuzzer's capability of analyzing complex smart contracts.

\begin{figure}[t]
	\centering
	\graphicspath{{figures/plots}}
	\def\svgwidth{0.9\linewidth}
	\begin{footnotesize}
		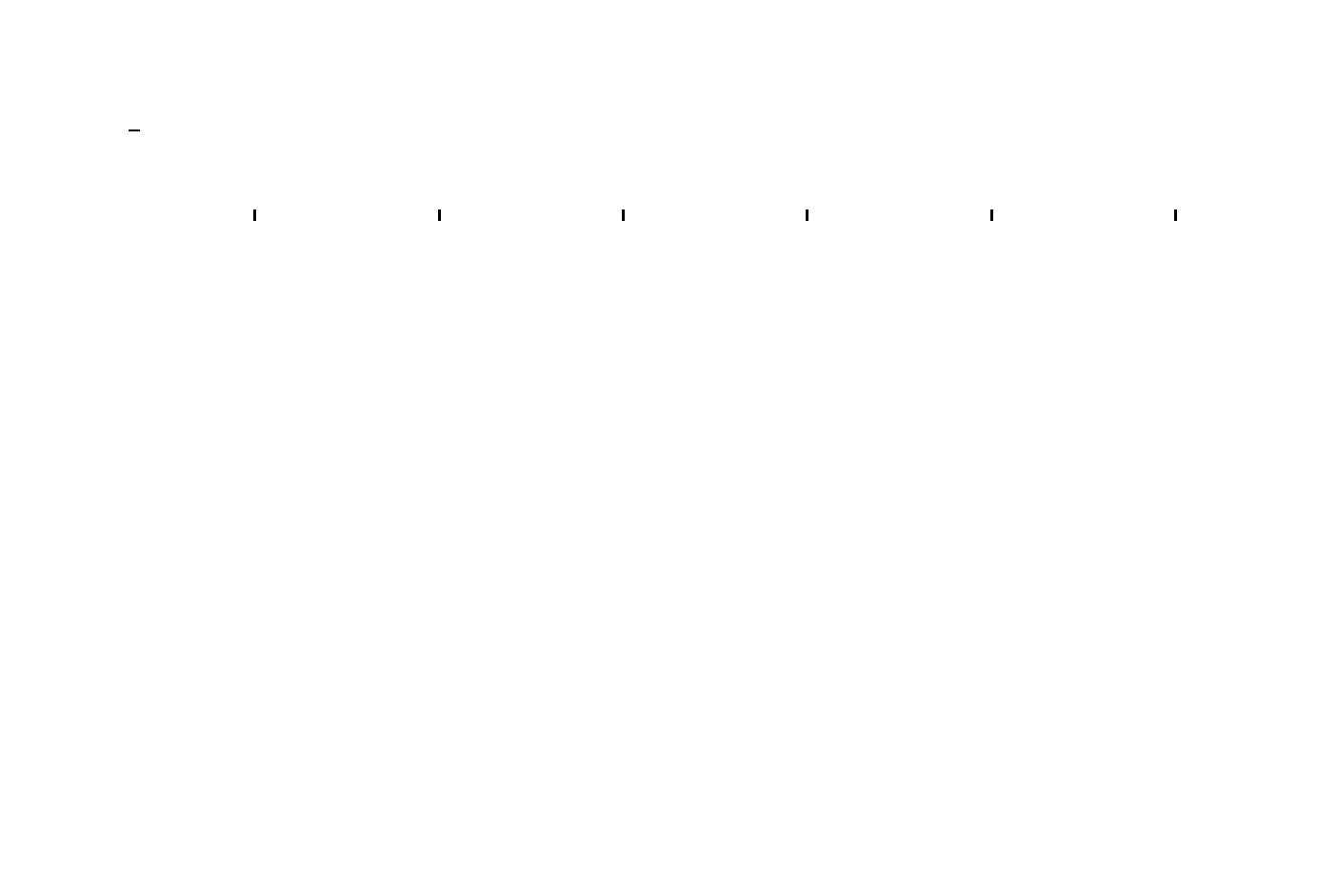
	\end{footnotesize}
	\caption{Increasing trend in smart contract complexity. We measure the complexity of all unique contracts with verified source code that appear in Ethereum until July 22, 2022.}%
	\label{fig:contractcomplexity}
	\vspace{-1em}
\end{figure}

\paragraph{Support for Complex Smart Contracts}
Owing to their increasing prevalence and ability to encode complex business logic, smart contracts are becoming increasingly more complex.
In \Cref{fig:contractcomplexity}, we measure how the complexity of deployed contracts has evolved over time.
We analyzed \num{120556} unique contracts with verified source code that appeared in the Ethereum blockchain until July 22, 2022.
For each contract, we measure the number of logical lines of code, the number of state-changing public functions, the number of comparison operators, and the number of branches in the control flow.
Our analysis confirms that contracts are becoming increasingly complex in all aspects.
For instance, among the unique smart contracts created in 2022, we find an average of roughly $530$ lines of code and $16$ state-changing public functions, while the same metric was at $182$ lines of code and $8$ functions in 2017.

To exercise all code paths while testing increasingly complex smart contracts, multiple consecutive transactions are required to explore the internal states of a smart contract.
However, most prior studies are only limited to rather short transaction sequences of length 3~\cite{KruppR18teether,NikolicKSSH18maian,FrankAH20ethbmc} or do not assess the ability to work with longer transaction sequences~\cite{Torres2021confuzzius,HeBATV19ilf,GriecoSCFG20echidna,Wustholz2020harvey}.
Typically, more complex contracts also require longer transaction sequences to cover different states of the contract during testing.
To assess the ability of existing analysis tools to cope with longer consecutive transaction sequences, we conducted an experiment with a set of benchmark contracts with artificial bugs.
Our experiment, summarized in \Cref{tab:bugbenchresults} and detailed in \Cref{sec:eval}, shows that existing analysis tools are not sufficient to analyze more complex contracts.
In particular, current fuzzing-based analysis tools~\cite{Torres2021confuzzius,GriecoSCFG20echidna} were unable to identify bugs that require a specifically ordered sequence of six or more transactions.
While symbolic execution tools~\cite{MossbergMHGGFBD19manticore,KruppR18teether,NikolicKSSH18maian} are capable of producing such sequences even up to ten transactions, they fail to identify faults that require accumulation of internal state over multiple transactions.
Apart from this, none of the analysis tools we tested are able to identify all bugs within a generous time budget of \num{48} hours.

\paragraph{Support for Complex Smart Contract Interactions}
Another challenge that we need to tackle is the frequent interaction of smart contracts with each other.
To precisely model such an interaction, whenever a smart contract calls another (potentially untrusted) smart contract, we must assume that the smart contract under test can be reentered at any function.
So-called reentrancy attacks have had devastating consequences in the past~\cite{dao-attack}.
To faithfully emulate the attacker's capabilities with respect to reentrancy, we must simulate the following scenarios:
At each call to an untrusted and potentially attacker-controlled contract, the target smart contract can be
\begin{inparaenum}[(1)]
	\item reentered at the same call depth multiple times,
	\item reentered at multiple functions, and
	\item reentered by a call originating from a different smart contract.
\end{inparaenum}
Current analysis tools mostly refrain from modelling arbitrary reentrant transactions due to state explosion~\cite{KalraGDS18zeus}; instead, most tools utilize over-approximative detectors for reentrancy bugs (i.e., state updates after calls~\cite{Tsankov2018securify, FeistGG19slither, Torres2021confuzzius}).

\begin{figure}[t]
	\inputminted{solidity}{figures/ReentrancyVulnBankBuggyLockHard.sol}
	\begin{center}
		\includegraphics[width=0.8\linewidth]{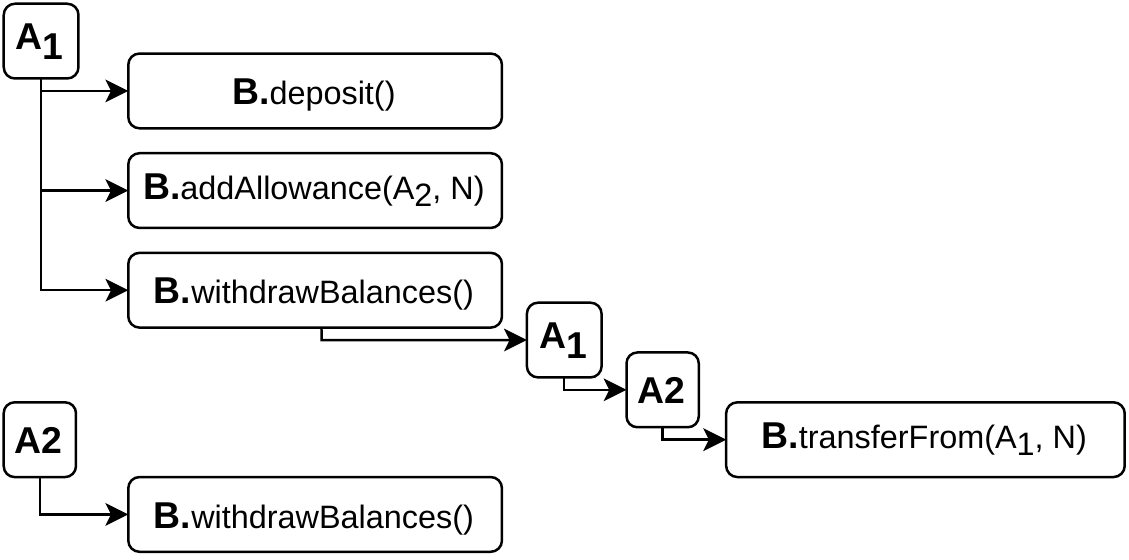}
	\end{center}
	\caption{A contract (Bank, $B$) with bypassable reentrancy-locking. The attack, depicted below the source code, requires two colluding attacker-controlled smart contracts ($A_1$, $A_2$).}
	\label{fig:hardre}
	\vspace{-1em}
\end{figure}

To better illustrate the challenge, consider the example in \Cref{fig:hardre}, which depicts a token-like contract with standard transfer and allowance mechanisms that is vulnerable to a reentrancy attack.
If using the \emph{checks-effects-interactions} code pattern~\cite{solidity-checks-effects-interaction} is not possible, the second best alternative to prevent reentrancy attacks is to use locking mechanisms~\cite{solidty-reguard}.
However, many analysis tools~\cite{FeistGG19slither,Tsankov2018securify,Torres2021confuzzius} do not handle locking mechanisms appropriately and simply report a potential reentrancy issue in the \emph{withdrawBalance} function for the locking mechanism itself and the balance update.
In the example in \Cref{fig:hardre}, the modifier \emph{withdrawAllowed} prevents an attacker from reentering the \emph{withdrawBalance} function.
This gives the developer a false sense of security, thinking that the \emph{userBalances} variables are protected by the locking mechanism and thus the contract must be secured against reentrancy attacks.
However, this assumes that an attack follows the call chain $A_1 \rightarrow B \rightarrow A_1 \rightarrow B$.
Since the attacker $A_1$ has arbitrary control, they can transfer control to a different colluding smart contract $A_2$, which allows to execute the call chain $A_1 \rightarrow B \rightarrow A_1 \rightarrow A_2 \rightarrow B$.
Reentrant calls from the second contract $A_2$ are not locked as it has not yet interacted with the target contract $B$.
Therefore, the second attacker contract $A_2$ can call into the \emph{transferFrom} function to move away the balance of $A_1$ before the call to \emph{withdrawBalance} finishes and resets the balance.
Using this attack, it is possible to bypass the reentrancy locking mechanism and withdraw twice the balance that was initially invested.

\section{Design of \toolnameslash}%
\label{sec:design}

\begin{figure*}[ht]
	\centering
	\begin{center}
		\includegraphics[width=0.90\textwidth]{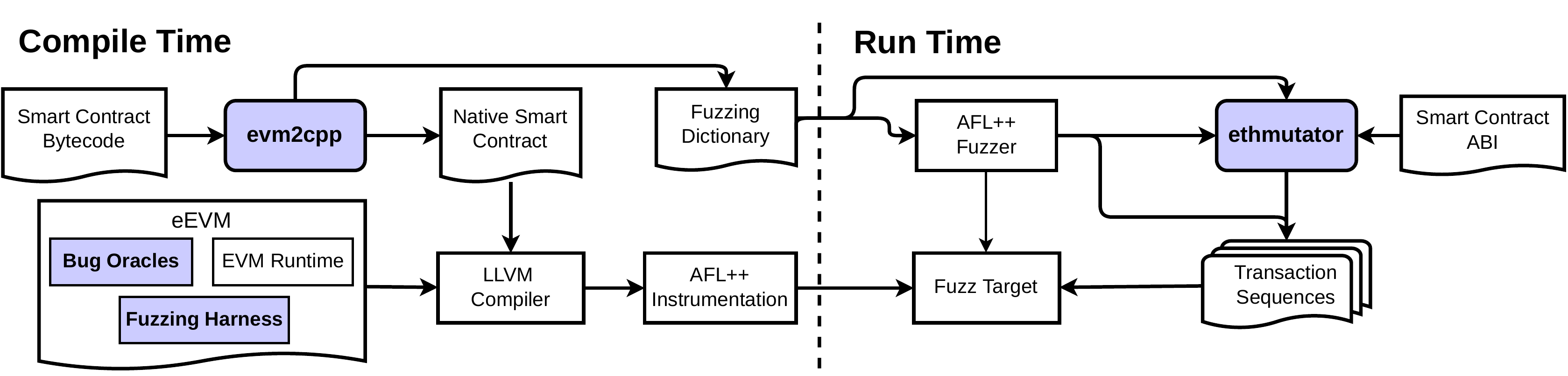}
	\end{center}
	\vspace{-1em}
	\caption{Architecture of \toolname.}
	\label{fig:architecture}
	\vspace{-1em}
\end{figure*}

The design of our \toolnamelong (\toolname) is driven by two major features: optimizing test case throughput and accurately modelling complex interactions with smart contracts.
To achieve the former, \toolname uses two explicit phases, a compile and a run time phase (see \Cref{fig:architecture}).
At compile time, the EVM bytecode of the smart contract is translated to \Cpp code with our newly developed \emph{evm2cpp} compiler and paired with a fuzzing-optimized EVM runtime, facilitating fast smart contract execution.
To accurately model interactions with smart contracts, we devise an approach to allow the fuzzer to mutate the behavior of multiple simulated attacker-controlled smart contracts.
Each generated test case specifies a sequence of transactions, which are executed by the fuzzing harness.
However, in contrast to prior work~\cite{KruppR18teether, GriecoSCFG20echidna, NikolicKSSH18maian,FrankAH20ethbmc, Torres2021confuzzius}, this transaction sequence also specifies the behavior of callbacks to attacker accounts, including return values and further reentrant transactions.
To detect bugs, \toolname features detectors that are directly built into the EVM runtime and the fuzzing harness.
Here \toolname supports a commonly featured Ether-based bug oracle that attempts to gain Ether, but also custom bug oracles that are specified by a developer in Solidity code.
In what follows, we discuss and explain our design choices in more detail.

\subsection{Modelling Blockchain Interaction}

To faithfully model complex (possibly adversarial) interactions on the blockchain, we define an input format for Ethereum transaction sequences that supports
\begin{inparaenum}[(a)]
	\item fuzzing the blockchain environment,
	\item fuzzing return data,
	\item reentrant transactions, and
	\item targeting multiple contracts.
\end{inparaenum}

\paragraph{Blockchain Environment}
\toolname runs the smart contract in a custom blockchain environment, which contains several attacker-controlled accounts.
A user of \toolname can also supply a custom initial blockchain state, e.g., to fuzz smart contracts that rely on other smart contracts or expect to be deployed at a certain address.
\toolname allows the fuzzer to choose and mutate several environmental values of the Ethereum environment.
For example, the fuzzer chooses the block number and timestamp at the beginning of the transaction sequence and is allowed to advance both at every transaction.
This enables us to handle smart contracts that expect a certain timespan to pass between two consecutive transactions.
Furthermore, the fuzzer can increase the initial Ether balance of the target contract to simulate prior Ether investment into the contract.

The blockchain state is reset before every executed test case, which ensures that each generated transaction sequence can be deterministically executed.
This is necessary to obtain reliable coverage measurements and eases root-cause analysis since the developer can reliably replay a transaction sequence.
In many cases, the transaction sequence can be directly utilized as an end-to-end exploit against the deployed version of the contract.

\paragraph{Transaction Sequence}
Every test case in \toolname consists of a header specifying the initial environment followed by a sequence of transactions.
Similar to regular Ethereum transactions, each transaction consists of a sender, a receiver, a call value (i.e., transferred Ether), and associated input data.
However, for performance reasons, we restrict the senders and receivers to a small set of accounts that are set when the fuzzer is launched.
In a typical single-contract fuzzing setup, the set of receivers will include only the target smart contract.

\toolname simulates the behavior of arbitrary smart contracts at the attacker-controlled accounts.
To achieve this, each transaction requires additional associated data beyond what a regular Ethereum transaction requires.
This includes fields that specify what to do if the target smart contract calls back to an attacker-controlled address.
Each transaction can have multiple associated \emph{return-headers}, which specify
\begin{inparaenum}[(1)]
	\item whether the call succeeded,
	\item what data to return,
	\item and how many reentrant calls can be performed.
\end{inparaenum}
The fuzzer is then free to choose arbitrary values for any of these parameters.
However, we bound the number of return headers per transaction to 255 and we also bound the number of reentrant transactions to 255 in our implementation of \toolname.
If \toolname encounters a callback without an associated \emph{return-header}, \toolname will simulate a failed call.
This allows the fuzzing process to explore a large variety of behaviors of attacker-controlled smart contracts.

\paragraph{Reentrant Transactions}
Current dynamic analysis tools~\cite{KruppR18teether,Jiang18contractfuzzer,Torres2021confuzzius} focus on generating lists of top-level transactions that trigger an exploit.
In contrast, \toolname simulates the behavior of a reentrancy-capable attacker.
We model the transaction sequences as a tree of transactions.
This enables \toolname to explore various shapes of the tree:
reentering the same function repeatedly, reentering the same function only once, reentering the same contract in a different function, or reentering the same contract multiple times at the same call-depth.

However, in practice, not all shapes of the tree are possible for a certain contract.
Only some of the functions of a contract allow for callbacks to the attacker and therefore further reentrant transactions.
In general, it is not possible to compute the shape of the tree in advance for all contracts.
Whether an external call to an attacker is performed by the target smart contract generally depends on the input and as such, cannot be determined before executing the transaction.

\toolname's fuzzing harness dynamically builds a transaction tree.
All other components of \toolname still operate on a list of transactions.
The harness uses the list of transaction as a queue: when an external call is encountered, the fuzzer can simulate a reentrancy with the next transaction in the queue.
If the transaction queue is empty no reentrant call is performed.
To mutate this ad-hoc tree structure, the fuzzer can mutate a single field in the return header that specifies how many reentrant transactions can be executed when an external call is encountered.
The fuzzing harness ignores this field if the transaction does not trigger an external call.

\begin{figure*}[t]
	\centering
	\includegraphics[width=0.98\linewidth]{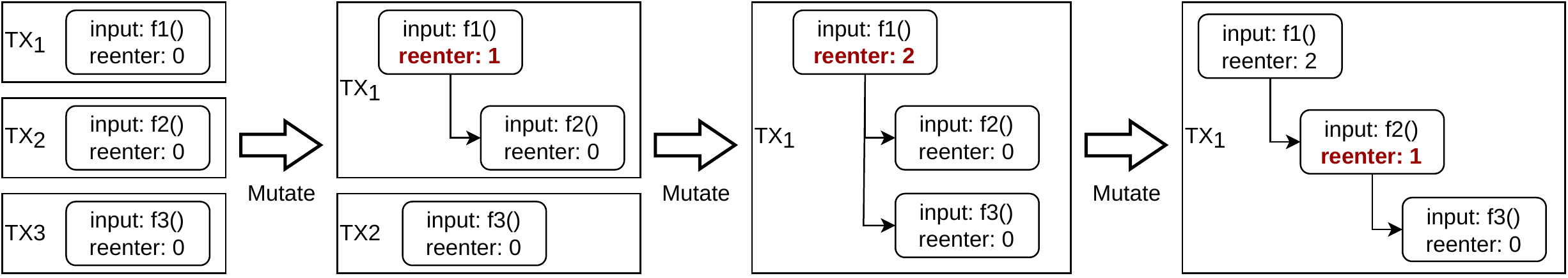}
  \caption{Mutating a transaction sequence to obtain reentrant transaction sequences with different shapes. The mutated \emph{reenter} field is highlighted in bold red.}%
	\label{fig:remut}
	\vspace{-1em}
\end{figure*}

We show an example of chain of mutations and how they affect the shape of a transaction tree in \Cref{fig:remut}.
We start with a flat sequence of three transactions that target the functions \emph{f1}, \emph{f2}, and \emph{f3}.
All transactions have the \emph{reenter} flag set to $0$.
The fuzzer then probabilistically mutates the \emph{reenter} flag to modify the shape of the transaction tree.
In the first transaction it is set to $1$.
If the call to \emph{f1} triggers a callback to the attacker, \toolname will perform a reentrant call with the second transaction in the queue, i.e., a reentrant call to \emph{f2}.
The third transaction remains a top-level transaction.
With this simple mutation, \toolname has now generated a cross-function reentrancy attack.
Setting the \emph{reenter} flag to $2$ the produces a tree with two reentrant transactions at the same call depth.
However, if the \emph{reenter} flag on the second transaction is set to a positive value, the third transaction becomes a reentrant transaction at a deeper call depth.
Since we do not have any more transactions in the queue, there is no second reentrant transaction at call depth \num{1}.

This approach to defining the transaction tree in an ad-hoc manner can model arbitrary tree shapes.
However, we introduce several limits in our implementation of \toolname because the shape of the transaction tree is bounded by several factors in practice.
Ethereum's gas mechanism limits the number of possible callbacks per transaction.
Therefore, in our implementation of \toolname, we limit the maximum number of callbacks to \num{255} and the number of reentrant transactions per callback to \num{255}.
Similarly, we stop execution when reaching the EVM defined maximum call depth of 1024.

\paragraph{Compositional Security}
Modern smart contracts are increasingly coupled with other smart contracts.
For example, token contracts are often tied to exchange contracts, where the token can be traded for other tokens or Ether.
Beyond the security of contracts in isolation, also the \emph{composition of multiple contracts} must remain secure against attacks (compositional security).
Recently, several attacks have been reported that were only possible due to composition of multiple smart contracts that were developed independently~\cite{Cecchetti2021serif,Torres21eyeofhorus,creamfinance,revestfinance}.
For example, the \emph{Uniswap} reentrancy attack was only possible because the Uniswap contract was combined with a new type of token contract that would perform a callback to the attacker.
The Uniswap contract did not expect reentrancy to be possible on an external call and is indeed safe when paired with most token contracts.

\begin{figure*}[ht]
	\centering
	\begin{subfigure}{0.24\linewidth}
		\includegraphics[width=\textwidth]{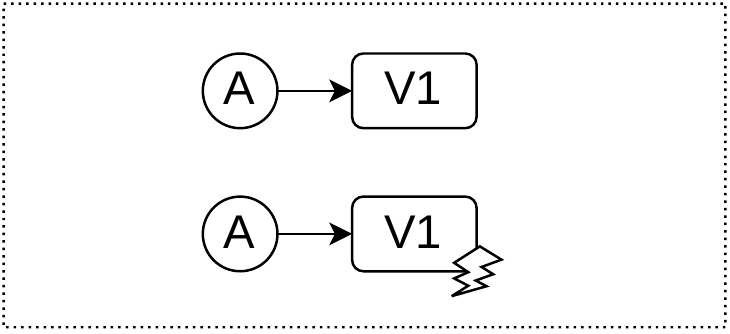}
		\caption{Single Contract.}%
		\label{fig:txseqs:normal}
	\end{subfigure}
	\hfill
	\begin{subfigure}{0.24\linewidth}
		\includegraphics[width=\textwidth]{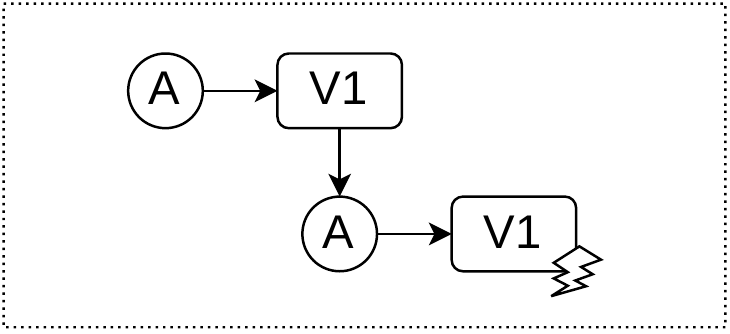}
		\caption{Single Contract Reentrancy.}%
		\label{fig:txseqs:reenter}
	\end{subfigure}
	\hfill
	\begin{subfigure}{0.24\linewidth}
		\includegraphics[width=\textwidth]{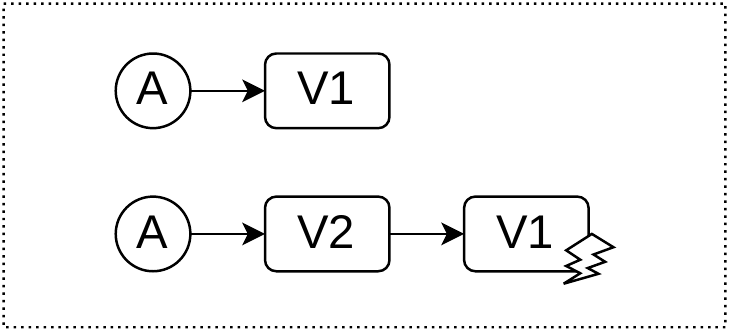}
		\caption{Contract Composition.}%
		\label{fig:txseqs:compositional}
	\end{subfigure}
	\hfill
	\begin{subfigure}{0.24\linewidth}
		\includegraphics[width=\textwidth]{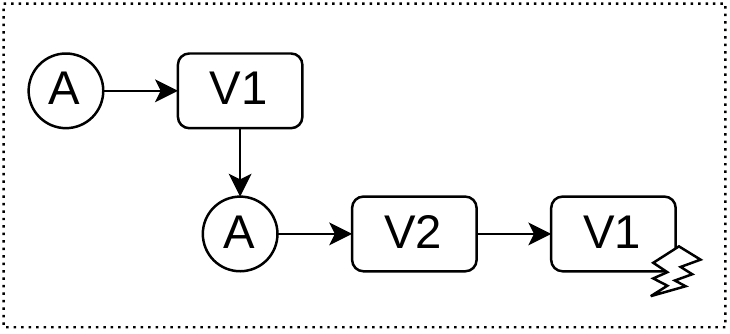}
		\caption{Compositional Reentrancy.}%
		\label{fig:txseqs:compositionalreenter}
	\end{subfigure}

	\caption{Different settings illustrating the difference and similarities between reentrancy and compositional attacks.}%
	\label{fig:reentrancyandcomposition}
	\vspace{-1em}
\end{figure*}

The most prominent examples of compositional security issues were also reentrancy attacks~\cite{Cecchetti2021serif}.
However, compositional attacks are not necessarily reentrancy attacks.
\Cref{fig:reentrancyandcomposition} shows four different attack settings, where the last (sub-)transaction triggers a bug in contract \emph{V1}.
\Cref{fig:txseqs:normal} depicts a flat sequence of transactions: two subsequent calls to contract \emph{V1}.
\Cref{fig:txseqs:reenter} shows a simple reentrancy attack, \emph{V1} call back into the attacker \emph{A}, which performs a reentrant call to \emph{V1}.
\Cref{fig:txseqs:compositional} depicts a compositional attack, where a composition of two contracts is vulnerable.
The DoS attack against the \emph{Parity Multisig Wallet} is an example of such an attack~\cite{parity-postmorterm}.
The bug can be triggered by setting up a vulnerable state and then forcing \emph{V2} to call \emph{V1}.
\Cref{fig:txseqs:compositionalreenter} depicts a compositional reentrancy attack, where the call from \emph{V2} to \emph{V1} is a reentrant call.
The \emph{Uniswap} attack is an example of such a compositional reentrancy bug~\cite{Cecchetti2021serif}.

\toolname's design also allows for testing compositions of smart contracts.
First, \toolname is designed such that it can import parts of the blockchain state to set up a realistic environment for the target smart contracts.
For example, developers could test the security of their deployed contract compositions by setting up their initial state on a local test chain and then running \toolname to detect potential issues.
Second, \toolname supports selecting the receiver of a transaction, including reentrant transactions.
A developer or analyst just needs to configure the set of contracts to be analyzed by \toolname in composition.

\subsection{Optimizing Test Case Throughput}

Most state-of-the-art analysis tools~\cite{GriecoSCFG20echidna, MossbergMHGGFBD19manticore, KruppR18teether, NikolicKSSH18maian, FrankAH20ethbmc, Torres2021confuzzius} develop or utilize custom EVM implementations that offer the introspection and extension possibilities necessary to perform dynamic smart contract analysis.
Similarly, as part of our high throughput fuzzing framework, we develop an execution environment for smart contracts that is optimized for fuzzing.
Here, test case throughput, i.e., both fast mutation/generation and fast execution, emerge as one of the most important properties of a fuzzer to achieve good results in practice.

\paragraph{Translating EVM to \Cpp}
Most widely used Ethereum clients implement an interpreter to execute EVM smart contracts.
Typically, Ethereum nodes execute a large number of different smart contracts throughout their operation; in this case, it suffices to rely on an interpreter.
However, in a fuzzing setting, the same smart contract is repeatedly executed.
As a consequence, the overhead of the interpreter adds up over time and causes significant overhead over longer fuzzing periods.
In \toolname, we remove this overhead by performing additional translation and optimization before starting the fuzzing campaign.
To remove the interpreter overhead, we develop \emph{evm2cpp}, a custom translation layer from EVM bytecode to \Cpp, and pair this with a customized EVM runtime optimized for fuzzing.
While this approach gives up the flexibility of an interpreter or a just-in-time compiler, an ahead-of-time compilation approach can utilize the full set of optimization techniques in modern \Cpp compilers, such as link-time optimization.
Furthermore, \emph{evm2cpp} performs optimizations to remove costly EVM stack operations.
Note that ahead-of-time translation of bytecode targeting a virtual machine to \Cpp~\cite{il2cppunity,Scholz2016souffle} or directly to native code~\cite{androidart,Wimmer2019graalnative} has been previously applied in various academic and industry settings to speed up execution time.
In \toolname, we apply this technique to EVM bytecode for the first time.

We discuss correctness of our translation approach in \Cref{sec:implementation:translation}.
During all of our fuzzing campaigns and experiments, we have not identified any bug or crash that was caused by \emph{evm2cpp} miscompilation.
Furthermore, to quantify the performance improvements of translation we conduct an ablation study (see \Cref{sec:eval:throughput}).

\paragraph{Optimizing the EVM runtime}
The smart contract that is translated to \Cpp still requires an EVM runtime that implements the opcode handlers and interaction with the blockchain.
We pair the \Cpp code generated by \emph{evm2cpp} with an EVM runtime, which we adapted and optimized from the \emph{eEVM} project~\cite{eEVM}.
We chose the \emph{eEVM} project because of its relatively simple codebase that can be easily extended and adapted for fuzzing.
This includes omitting or simplifying several features required by a full EVM implementation operating as part of an Ethereum node.
For example, our EVM runtime does not feature instruction-accurate gas tracking.
We do not need the gas mechanism to limit execution time, since it is limited per test case by the fuzzer.
Detecting the majority of vulnerabilities, such as access control or reentrancy, does not require instruction-accurate gas tracking.
However, checking the gas budget is necessary to accurately execute external calls.

Furthermore, we stop test case execution at the first failing transaction.
Since the failing transaction would be rolled-back, this has no effect on subsequent transaction executions.
Instead of performing roll-backs after every failing transaction, we stop execution of the current test case, reset the state of the Ethereum blockchain to its initial state, and let the fuzzer generate a new test case.
This approach nudges the fuzzer to generate transaction sequences that only contain succeeding transactions.
The only exception is the last transaction in the sequence, which is allowed to explore error handling paths.
Furthermore, this approach increases the effectiveness of test case splicing: When \toolname combines two previously generated test cases into one, it will very likely generate a new test case containing only succeeding transactions.

\paragraph{Optimizing the Mutations}
At run time, the base fuzzer launches the target smart contract with seed inputs, i.e., a seed with a single transaction without any input.
The fuzzing process is driven by a greybox fuzzing approach that involves mutating inputs, executing the fuzz target, and measuring the code coverage to find new interesting transaction sequences.
Note that the base fuzzer is unaware of the structure that is inherent to the test cases, i.e., the structure of the transaction sequence and the structure of the individual transaction inputs.
The base fuzzer's mutation strategies are not efficient in mutating the structure of the input.
Hence, we augment the base fuzzer with a carefully engineered and optimized test case generator and mutator that performs structure-aware mutations and generations.
Our custom mutator is called \emph{ethmutator} and performs both
\begin{inparaenum}[(1)]
	\item mutations on the transaction inputs according to the smart contract's ABI and
	\item structural mutations on the transaction sequence.
\end{inparaenum}

\subsection{Bug Oracles}

Prior work on smart contract fuzzing attempted to address two orthogonal problems at the same time: triggering bugs and detecting bugs~\cite{NguyenP0L020sfuzz,HeBATV19ilf,Jiang18contractfuzzer,Torres2021confuzzius,Choi2021smartian}.
Bug oracles are dynamic analyses that signal the fuzzer that a fault was triggered.
In this paper, we focus primarily on the aspect of triggering faults by developing a high throughput fuzzer to identify the right input to trigger a fault.
We opt to primarily use a simple---yet powerful---bug oracle: Ether gains.
This is in line with prior work on exploit generation for smart contracts~\cite{KruppR18teether,NikolicKSSH18maian,FrankAH20ethbmc}.
However, we also support custom bug oracles defined in Solidity code by the smart contract developer to cover contract-specific bug classes.

\toolname defines an attack to satisfy one of the following conditions:
\begin{inparaenum}[(a)]
	\item The attacker is able to trigger a \emph{selfdestruct} to an arbitrary address, thereby allowing the attacker to drain the funds of the contract and create a \emph{Denial-of-Service} scenario.
	\item The attacker can redirect the control-flow to an arbitrary address (using the \texttt{DELEGATECALL} instruction), which would give the attacker control over the target's Ether.
	\item The attacker is able to gain Ether by interacting with the contract, i.e., the sum of the balances of the attacker-controlled contracts must exceed the initial sum of balances of these contracts.
\end{inparaenum}
In contrast to other bug oracles, this approach avoids a high number of false alarms by design.
For example, accurately detecting integer overflows~\cite{Torres2018osiris} or reentrancy~\cite{sereum} without high-level type information is challenging.
However, it is comparably straightforward to detect if a fuzzer generates a transaction sequence exploiting an integer overflow or reentrancy to gain Ether.
Interestingly, we found that this type of bug oracle is also \emph{more accurate} than bug oracles implemented in other analysis tools.
For example, the contract depicted in \Cref{fig:leoracleproblem} is not identified as vulnerable by either of two state-of-the-art hybrid fuzzers, Confuzzius~\cite{ Torres2021confuzzius} and Smartian~\cite{Choi2021smartian}.
\toolname's Ether-gains bug oracle turns out to cover more cases such as this example.

However, an Ether-based bug oracle often does not capture the semantics of some smart contract applications, such as token contracts.
To tackle such contracts, we also implemented support for custom invariant checking and assertion checking.
Smart contract developers specify invariants that the fuzzer tries to invalidate.
We implement support for mechanisms that are also utilized in other analysis tools, such as Echidna~\cite{GriecoSCFG20echidna} and Mythril~\cite{mythril} (see also \Cref{sec:impldetails}).
Developers that already utilize one of these tools can directly reuse their custom invariants and assertions with \toolname.

\begin{figure}[t]
	\centering
	\inputminted{solidity}{figures/etherdrainother.sol}
	\caption{Contract that is not considered as vulnerable by the \emph{leaking Ether} bug oracles of Confuzzius~\cite{ Torres2021confuzzius} and Smartian~\cite{Choi2021smartian}, but detected by \toolname's Ether gain bug oracle.}%
	\label{fig:leoracleproblem}
	\vspace{-1em}
\end{figure}

\section{\toolnameslash Implementation}%
\label{sec:implementation}

We now present an overview of the implementation of \toolname. 
We release our implementation as open source at \releaseLink.
We discuss further minor technical details in \Cref{sec:impldetails}.

\subsection{EVM to \Cpp Translation}%
\label{sec:implementation:translation}

The \emph{evm2cpp} component is a custom compiler that translates EVM bytecode to \Cpp, implemented in roughly \num{2500} lines of Rust code.
First, we perform a linear pass over the EVM bytecode to build the set of all basic blocks.
In EVM bytecode, all jump destinations are marked with \texttt{JUMPDEST} instructions.
We use these to identify the boundaries of basic blocks by looking for branching instructions and jump destination markers.
We do not construct a full control-flow graph, avoiding costly and error-prone analysis.
Instead, we perform local analysis and optimization at the EVM basic block level.
We emulate basic blocks in isolation using abstract values as placeholders for non-constants to perform data-flow analysis and constant propagation with respect to the EVM stack.
Contrary to full abstract interpretation, we stop emulation at the end of a basic block and therefore do not need to handle control-flow instructions.

The code generation procedure translates each EVM basic block to a \Cpp lexical block.
If we can infer the jump target at the end of a basic block with our constant propagation, we directly emit a \emph{goto} statement to the target \Cpp lexical block.
Otherwise, we have to fall back to using a large jump table via the \emph{computed goto} feature.
In both cases the \Cpp goto is translated into a jump instruction by the \Cpp compiler.
The gotos between translated basic blocks are then instrumented by the coverage-instrumentation pass of the fuzzer.

Within a basic block, each opcode is translated to a call to the respective opcode handler function in the EVM runtime.
We translate the stack-based EVM opcodes into a lightweight register-based form, where each register is translated to a \Cpp local variable and no register is reused (similar to single-static assignment form, albeit without the need for the $\phi$ instruction).
At the beginning and the end of each translated basic block, we ensure that the stack effects of the register-based form and the original EVM opcodes are the same.
Essentially, we use the EVM stack exclusively to pass parameters between translated basic blocks.
This enables us to eliminate a number of costly EVM stack operations and emit \Cpp code that can be well optimized by recent clang versions (we tested \emph{clang}~$\geq 13$).

\Cref{fig:codegen} shows an example of a translated basic block.
Here, the \texttt{DUP1} opcode is eliminated, which duplicates a value on the EVM stack.
Furthermore, we eliminate the \texttt{PUSH2} instruction, which pushes a constant to the stack.
Owing to the constant propagation pass we perform, we can infer that this constant is used as a jump target later.
Instead of dispatching the jump via the EVM stack, we emit a \emph{goto} statement that directly targets the desired block.
Before the jump, we fix up the EVM stack effects of the basic block by pushing the right values to the stack.
With respect to the EVM stack, the original bytecode performs three pushes, two pops, and one replacement of the top element.
In contrast, the generated \Cpp code uses only a single stack push.

\begin{figure}[t]
	\begin{minted}{C}
pc_5a : {
  /* JUMPDEST */
  /* CALLVALUE */
  const uint256_t v_1_0 = callvalue_v();
  /* DUP1 */
  /* ISZERO */
  const uint256_t v_3_0 = iszero_v(v_1_0);
  /* PUSH2 0x66 */
  /* JUMPI */
  if (v_3_0) {
    ctxt->s.push(v_1_0);
    goto pc_66;
  }
  ctxt->s.push(v_1_0);
  goto pc_62;
}
\end{minted}
	\caption{Example for basic block translated by \emph{evm2cpp}. The comments show the original instructions; some stack-related opcodes have no direct counterpart in the emitted \Cpp code.}%
	\label{fig:codegen}
	\vspace{-1em}
\end{figure}

\paragraph{Performance Improvements}
The translation approach allows for two major optimizations that result in execution speed-ups:
\begin{inparaenum}[(1)]
  \item removal of the interpreter loop, and
  \item removal of EVM stack operations.
\end{inparaenum}
The former allows us to remove many repeated lookups of the opcode handlers and indirect dispatching.
Because there is no indirect dispatch to opcode handlers, the C++ compiler is now able to inline opcode handlers in the emitted native code to avoid the overheads of calls.
The EVM stack is typically implemented as a high-level data structure that resides on the native-code heap, e.g., a C++ \emph{vector}.
Accessing the EVM stack requires bounds and stack-overflow checking, resulting in significant overhead.
The optimizations in \emph{evm2cpp} remove many stack operations, resulting in faster execution.
We present an ablation study in \Cref{sec:eval:throughput} to quantify the performance improvements due to \emph{evm2cpp}.

\paragraph{Correctness}
Correctness of our compiler is paramount for \toolname.
Our compilation approach is designed with minimum error potential.
We utilize the EVM stack to pass data between translated basic blocks in the translated bytecode.
Optimizations are performed within a basic block, allowing us to avoid error-prone and costly indirect jump target analysis.
Nevertheless, our translation approach intentionally diverts from the EVM specification to enable optimizations.
For example, due to the removal of EVM stack operations during compilation, a translated contract could temporarily exceed the maximum stack size imposed by the EVM specification.
However, this does not raise any problems for \toolname as it must not be able to execute potentially malicious bytecode.

To detect inconsistencies early in development, we tested \emph{evm2cpp} using a differential fuzzing approach.
We execute both the original eEVM interpreter and \emph{evm2cpp} code and check whether they behave the same.
In all of our experiments (see \Cref{sec:eval}), we did not encounter any issues that were caused by a miscompilation by \emph{evm2cpp}.

\paragraph{EVM-level Auto-dictionary}
To increase fuzzing efficiency, many fuzzers scan the code for constants and create a dictionary of potentially interesting values (e.g., file format header magic values).
However, they typically scan for up to \SI{64}{\bit} constants or null-terminated C strings and thus do not properly handle the \SI{256}{\bit} EVM words or \SI{160}{\bit} Ethereum addresses.
Hence, we generate a dictionary of values based on the constants discovered during the constant propagation pass.
This includes computed quasi-constants that are often found in EVM bytecode.

\paragraph{Dynamic Contract Creation}
\toolname executes contract constructors in the interpreter once before the start of the fuzzing run and fully supports common patterns such as proxy contracts.
However, \toolname does not support fuzzing contracts that create other contracts at runtime, as \toolname would have to execute previously unknown EVM bytecode, which is not possible in the current ahead-of-time compilation model.
When \toolname encounters an instruction that creates a new contract, \toolname will stop executing the current transaction sequence.
However, we only encountered a single contract that creates a new contract at runtime during our evaluation.
We believe that these cases are sufficiently rare to leave exploring this as future work.

\subsection{Fuzzing Harness}

We opted for a lightweight EVM implementation as the base for our fuzzing-optimized EVM runtime.
To this end, we adapted the open-source \emph{eEVM} project~\cite{eEVM} such that it fits the code-generation of \emph{evm2cpp} and added an implementation of several newer EVM opcodes, missing features, and various minor fixes.

\paragraph{Input Format}
Within the \emph{eEVM} project, we created a \emph{libfuzzer}-compatible fuzzing harness.
The bug oracles are directly integrated into the fuzzing harness and runtime support code of the \emph{eEVM} project.
The fuzzing harness features a parser for a custom input format we developed.
This format can be quickly parsed without ever failing, i.e., any unneeded data is discarded by the harness; for any missing data fields, default values are assumed.
This robust parsing approach allows the use of standard bit-flipping mutations~\cite{Fioraldi2020aflpp} that are unaware of the input structure.
The input format consists of an initial header defining the initial environment, followed by a sequence of transactions.
Each transaction is represented as a header and the transaction input.
Mutating the header for a transaction allows the fuzzer to select transaction-specific parameters, such as the sender account, the call value, and the number of allowed reentrant transactions.
For the input data, the parser simply consumes bytes from the fuzzer-provided data according to the input length specified in the header until the end of the fuzzer-provided data.
\Cref{fig:hardeattack} in the Appendix shows an example for a test case generated by \toolname to exploit the contract depicted in \Cref{fig:hardre}.
We designed the input format such that it enables high throughput fuzzing, while being expressive enough to model complex smart contract interactions.
Furthermore, we use the input format as a template to synthesize Solidity attack smart contracts that exploit the target.
For each attacker-controlled account, we synthesize a Solidity contract that implements the behavior as specified by the generated test case.

\paragraph{Fuzzer and Harness Integration}
While the fuzzing harness itself is mostly oblivious to the used fuzzer, we opted to rely on AFL++~\cite{Fioraldi2020aflpp} as one of the most advanced general-purpose fuzzers.
AFL++ supports various modern fuzzing techniques, such as collision-free coverage bitmaps, a Redqueen~\cite{Aschermann2019redqueen} implementation called \emph{cmplog}, and support for custom mutators.
Due to our transpilation approach, we are able to directly leverage the instrumentation of AFL++ for smart contract code.
We built the fuzzing harness with \emph{clang}, with the highest optimization setting and link-time optimization (LTO) enabled, instrumenting the harness with AFL++'s LTO-based collision-free code coverage instrumentation.
Since we have translated the EVM bytecode to \Cpp code, we can utilize AFL++'s coverage instrumentation to instrument the combination of harness and transpiled smart contracts.

However, we noticed a problem with AFL++'s implementation of the Redqueen mutations~\cite{Aschermann2019redqueen}.
By default, the optimized big integer library used by \emph{eEVM} uses branchless code when comparing the four \SI{64}{\bit} words that make up a single \SI{256}{\bit} value.
AFL++'s \emph{cmplog} does not detect when only one of the four words matches, as no new code coverage can be observed.
As a consequence, it fails to incrementally solve comparisons with large constants.
However, this issue is only relevant for bypassing comparisons and not during other arithmetic operations.
We opted to manually adapt the relevant functions in the EVM codebase to provide explicit coverage feedback to AFL++.
This allows AFL++'s \emph{cmplog} to solve a considerable number of fuzzing roadblocks caused by integer comparisons.
To further increase fuzzing efficiency when applying structural mutations in the custom mutator, we added a lightweight tracing mode for certain opcodes (comparisons and returns) to the codebase.
This enables us to identify quasi-constants and add them to the dictionary of our custom mutator at runtime.

When fuzzing for reentrancy attacks, we found it beneficial to notify the fuzzer about the call depth of the current execution.
To this end we introduce a call-depth-sensitive coverage reporting in the fuzzing harness.
Whenever a new basic block is executed, we record the current call depth in AFL++'s coverage map.
This allows AFL++ to distinguish executions of the same contract at different call depths.
Since, AFL++ receives a new coverage signal when a transaction is executed in a reentrant manner, the test case will be stored in the queue.
In turn, this increases the probability of finding reentrancy attacks.

\subsection{Custom Mutator}

We implemented a mutator library, called \emph{ethmutator}, in roughly \SI{10}{\kilo\loc} of Rust to efficiently generate and mutate:
\begin{inparaenum}[(1)]
	\item the structured transaction input expected by the smart contract code, and
	\item the transaction sequence input format parsed by the fuzzing harness.
\end{inparaenum}
The mutator library features a parser and emitter for the binary input format accepted by the harness code.
Based on this library, we implement several related tools, such as a structured test case minimizer and an AFL++ custom mutator.
The mutator is carefully engineered with high performance in mind.
We reduce the number of required allocations and copy operations by applying copy-on-write semantics while performing mutations on a transaction sequence.
We also use \emph{mimalloc}~\cite{Leijen2019mimalloc} to increase the performance of the mutator by a factor of four.

In Ethereum, a transaction is associated with an input field, which is simply a variable-length byte string.
Smart contracts use a de-facto standardized ABI format, which specifies how function calls with parameters of complex types are encoded.
To enhance the efficiency, we use the ABI information in the \emph{ethmutator} to perform mutations according to the ABI.
Unlike existing general-purpose fuzzers, our custom mutator can handle the complexity of the ABI format by acting as a grammar fuzzer for the given ABI and generating structurally-valid inputs based on it.
When choosing the values for primitive types, we rely on a fuzzing dictionary built into the custom mutator.
Recall that this dictionary is seeded with the constants that are discovered during the analysis pass of \emph{evm2cpp}.
In addition, we extend the dictionary with ``interesting'' values that are likely to trigger bugs (e.g., the dictionary contains the maximum value for every integer type supported by Solidity to make it more likely to trigger integer overflows).
When no ABI is available, we exploit the fact that ABI-encoded data is always similarly structured for efficient mutations.
For example, when appending a new transaction, we first select a \SI{4}{\byte} constant from the dictionary as a prefix for the input.
Since the basic unit of the ABI is a \SI{256}{\bit} EVM word, most of the input mutations operate on this word size if no ABI is available.

Another task of the custom mutator is to apply structured mutations to the transaction sequence.
We implemented several mutations such as adding, duplicating, or dropping transactions.
Furthermore, we implemented more involved mutators such as test case splicing or value propagation between transactions in a sequence.
Whenever the base fuzzer adds a test case to its queue, the custom mutator parses this test case and keeps the transaction sequence in memory.
The structured splicing mutation then replaces a randomly selected transaction sub-sequence with a sub-sequence obtained from a previous test case.
The intuition here is that transaction sequences from prior test cases contain valid transaction combinations.
Combining two valid transaction sequences is more likely to result in a new valid transaction sequence.
We also propagate values from earlier transactions to later transactions.
Hence, with some probability, values in the transaction input will be replaced with values that occurred as parameters in the input of earlier transactions.
Similarly, we set the value of \emph{address} types in the ABI to the address of attacker-controlled accounts that previously already sent a transaction.
Similar to AFL~\cite{aflhomepage}, the custom mutator has multiple stages and a fixed set of mutations that is applied to every test case.
Subsequently, random mutations are applied (similar to the \emph{havoc} phase in AFL).
The custom mutator also uses stacked mutations, where different random mutation operations are combined.

\section{Evaluation}%
\label{sec:eval}

In this section, we evaluate various aspects of \toolname and compare them to the current state-of-the-art in fuzzing and symbolic execution.
We evaluate \toolname with respect to scalability to longer transaction sequences, the test case throughput, and its bug detection capabilities.

\subsection{Scalability Benchmarks}%
\label{sec:eval:scalability}

\begin{table}[b]
	\newcommand{\tyes}{\color{ForestGreen}{$\checkmark$}}
	\newcommand{\tno}{\color{red}{$\times$}}
	\newcommand{\tmay}{\tyes}
	\newcommand{\tna}{-}
	\newcommand{\tun}{?}
	\centering
	\vspace{-1em}
	\caption{Capability of analysis tools to identify bugs with increasing transaction sequence length. {\tyes} bug can be found, {\tno} bug never found within \SI{48}{\hour}. Type: \textsc{S} symbolic execution, \textsc{F} fuzzer, \textsc{H} hybrid fuzzer.}%
	\label{tab:bugbenchresults}
	\resizebox{\linewidth}{!}{%
		\begin{tabular}{@{}l|c|ccccc|ccc|cccc@{}}
			\toprule
			Tool                                                     & Type                        & \multicolumn{5}{c}{multi}                            %
			                                                         & \multicolumn{3}{c}{complex}                                                        %
			                                                         & \multicolumn{4}{c}{justlen}                                                        \\
			                                                         &                             & 2                         & 3-7   & 8     & 9 & 10 & %
			5                                                        & 7                           & 9                                                    %
			                                                         & 8                           & 64                        & 128   & 256              %
			\\
			\midrule
			teEther~\cite{KruppR18teether}                           & \textsc{S}                  &
			\tyes                                                    & \tyes                       & \tno                      & \tno  & \tno  &          %
			\tyes                                                    & \tno                        & \tno                      &                          %
			\tno                                                     & \tno                        & \tno                      & \tno                     %
			\\
			MAIAN~\cite{NikolicKSSH18maian}                          & \textsc{S}                  &                                                      %
			\tyes                                                    & \tyes                       & \tyes                     & \tyes & \tyes &          %
			\tyes                                                    & \tyes                       & \tno                      &                          %
			\tyes                                                    & \tno                        & \tno                      & \tno                     %
			\\
			EthBMC~\cite{FrankAH20ethbmc}                            & \textsc{S}                  &                                                      %
			\tyes                                                    & \tno                        & \tno                      & \tno  & \tno  &          %
			\tno                                                     & \tno                        & \tno                      &                          %
			\tyes                                                    & \tno                        & \tno                      & \tno                     %
			\\
			Manticore~\cite{MossbergMHGGFBD19manticore}              & \textsc{S}                  &                                                      %
			\tyes                                                    & \tyes                       & \tno                      & \tno  & \tno  &          %
			\tno                                                     & \tno                        & \tno                      &                          %
			\tno                                                     & \tno                        & \tno                      & \tno                     %
			\\
			ConFuzzius~\cite{Torres2021confuzzius}                   & \textsc{H}                  &                                                      %
			\tyes                                                    & \tyes                       & \tno                      & \tno  & \tno  &          %
			\tno                                                     & \tno                        & \tno                      &                          %
			\tyes                                                    & \tyes                       & \tyes                     & \tmay                    %
			\\
			Echidna~\cite{GriecoSCFG20echidna}                       & \textsc{F}                  &                                                      %
			\tyes                                                    & \tyes                       & \tmay                     & \tno  & \tno  &          %
			\tno                                                     & \tno                        & \tno                      &                          %
			\tyes                                                    & \tyes                       & \tyes                     & \tyes                    %
			\\
			VeriSmart~\cite{So2021verismartsmartest} & \textsc{S}                  &                                                      %
			\tyes                                                    & \tyes                       & \tyes                     & \tyes & \tyes &          %
			\tyes                                                    & \tyes                       & \tno                      &                          %
			\tyes                                                    & \tno                        & \tno                      & \tno                     %
			\\
			Smartian~\cite{Choi2021smartian}                         & \textsc{H}                  &                                                      %
			\tyes                                                    & \tyes                       & \tno                      & \tno  & \tno  &          %
			\tno                                                     & \tno                        & \tno                      &                          %
			\tyes                                                    & \tyes                       & \tyes                     & \tyes                    %
			\\
			\midrule
			\toolname                                                & \textsc{F}                  &                                                      %
			\tyes                                                    & \tyes                       & \tyes                     & \tmay & \tyes &          %
			\tyes                                                    & \tmay                       & \tmay                     &                          %
			\tyes                                                    & \tyes                       & \tyes                     & \tyes                    %
			\\
			\bottomrule
		\end{tabular}
	}

	\let\tyes\undefined
	\let\tno\undefined
	\let\tna\undefined
	\let\tun\undefined
	\let\tmay\undefined
\end{table}

\begin{figure*}[t]
	\centering
	\begin{subfigure}[b]{0.31\linewidth}  %
		\graphicspath{{figures/}}
		\def\svgwidth{\linewidth}
		\begin{scriptsize}
		    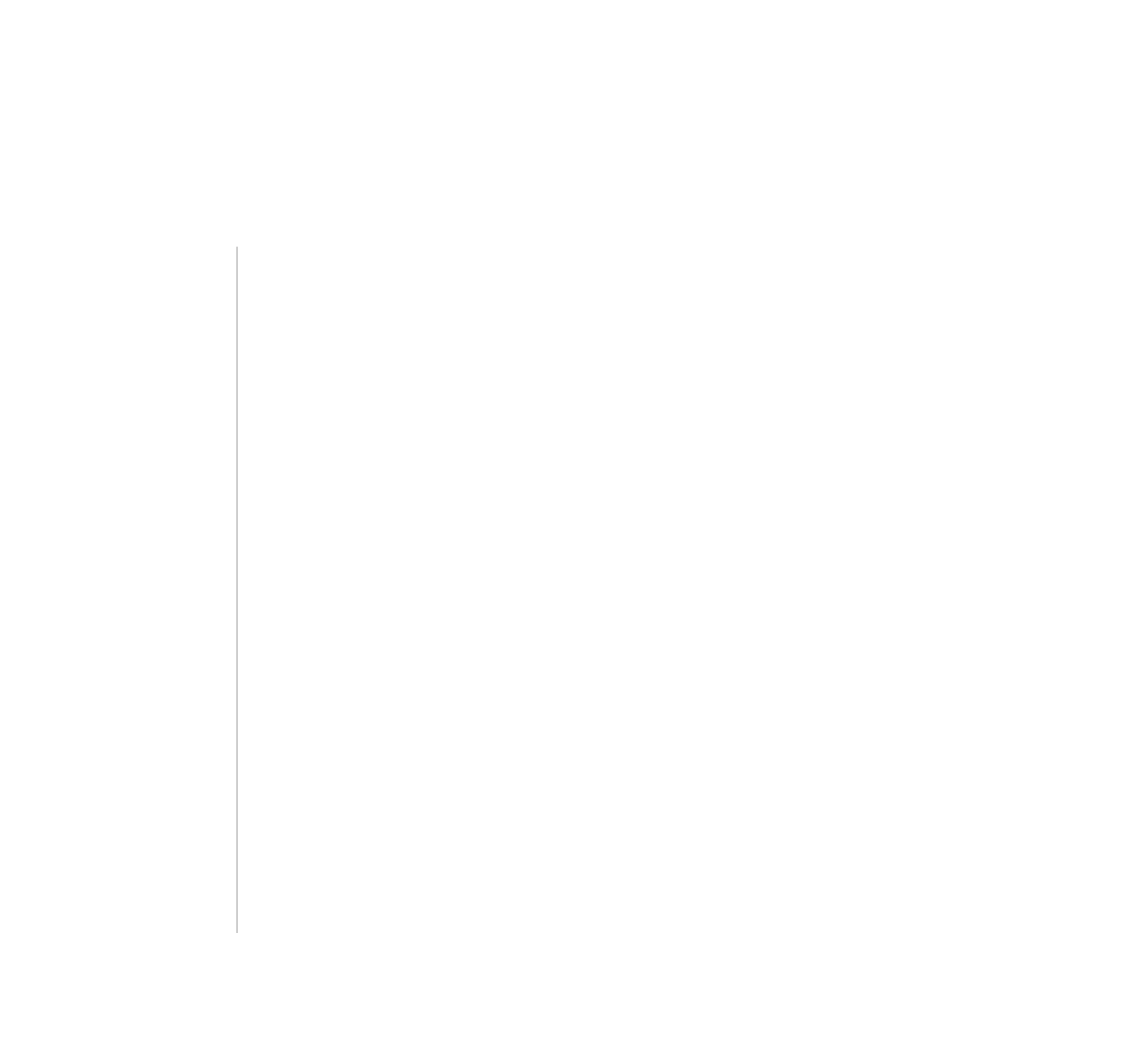
		\end{scriptsize}
		\vspace{-1em}
		\caption{\emph{multi}}%
		\label{fig:multigenplot}
	\end{subfigure}
	\hfill
	\begin{subfigure}[b]{0.31\linewidth}  %
		\graphicspath{{figures/}}
		\def\svgwidth{\linewidth}
		\begin{scriptsize}
			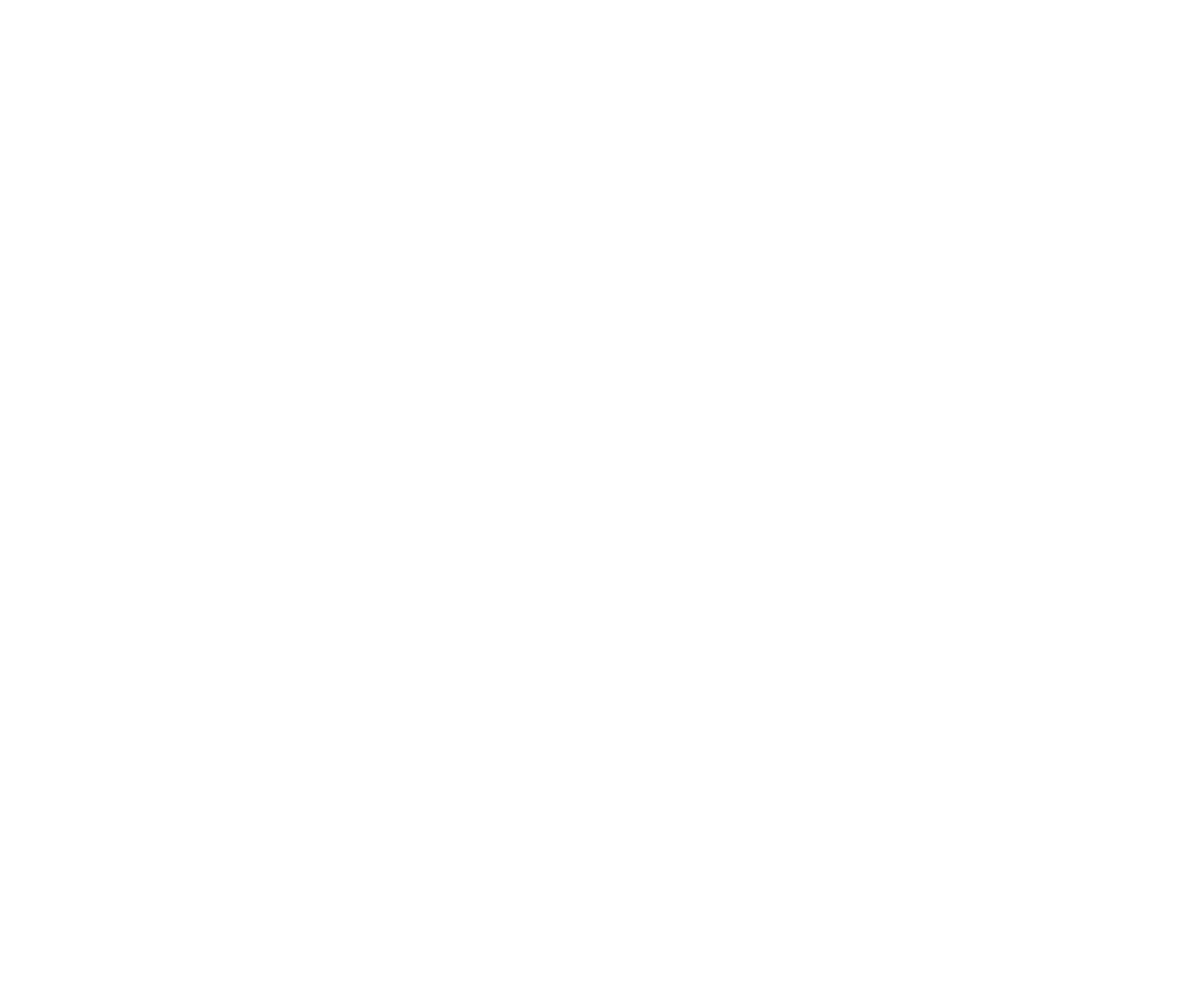
		\end{scriptsize}
		\vspace{-1em}
		\caption{\emph{complex}}
		\label{fig:sub:multiman}
	\end{subfigure}
	\hfill
	\begin{subfigure}[b]{0.31\linewidth}  %
		\graphicspath{{figures/}}
		\def\svgwidth{\linewidth}
		\begin{scriptsize}
			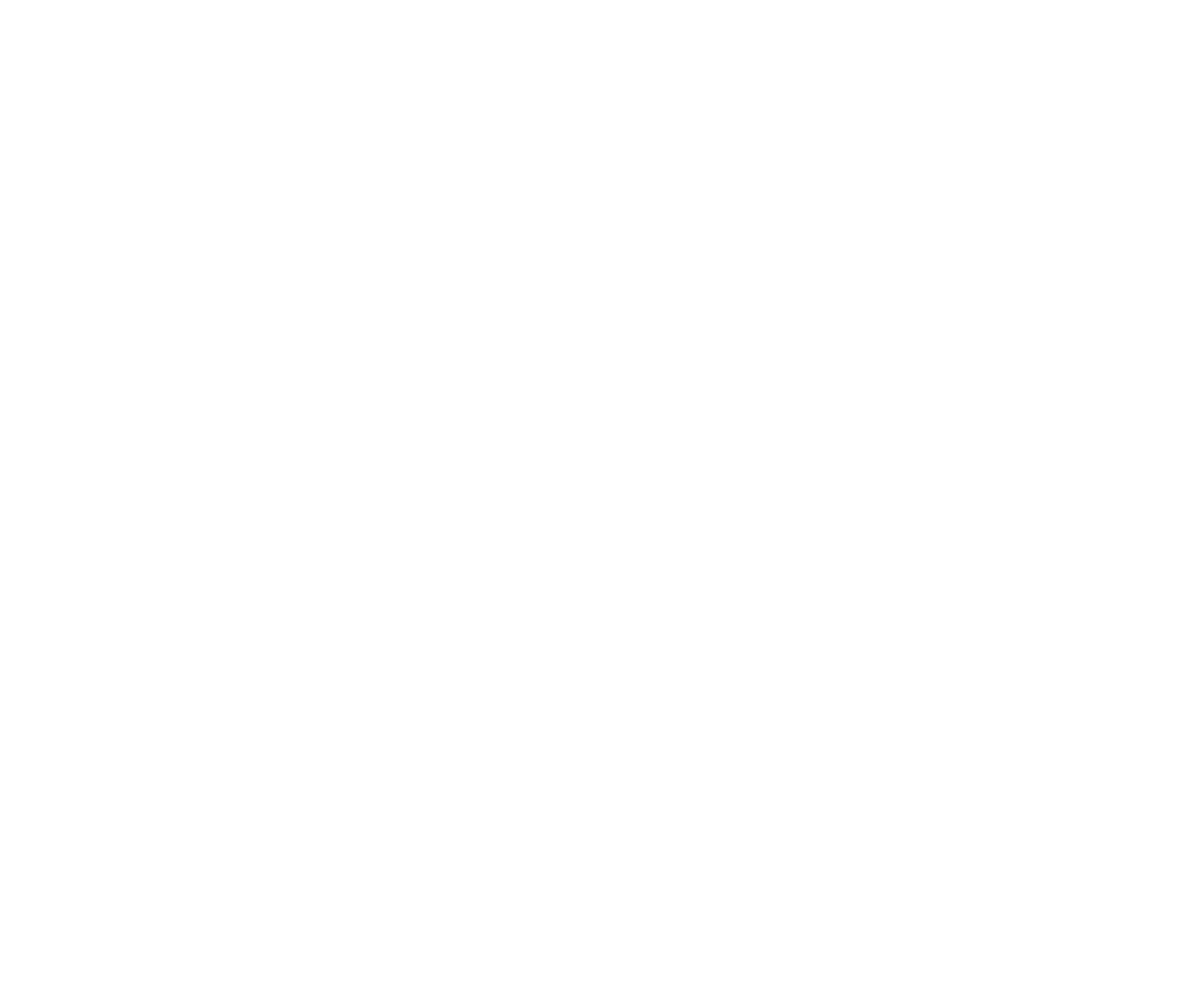
		\end{scriptsize}
		\vspace{-1em}
		\caption{\emph{justlen}}
		\label{fig:sub:justlen}
	\end{subfigure}
  \caption{Results of scalability experiments showing the analysis time required over the length of transaction sequences. Tools with the postfix \emph{c4} were run in parallel on 4 cores.}%
	\label{fig:otherplots}
	\vspace{-1em}
\end{figure*}

We start with an evaluation of the effectiveness of analysis tools in dealing with an increasing length of transaction sequences.
To this end, we created a benchmark consisting of three types of contracts (\emph{multi}, \emph{complex}, and \emph{justlen}), which model different code structures and roadblocks (that hinder analysis) typically found in smart contracts.
For each type of contract, we devise several variants (9 for \emph{multi}, 3 for \emph{complex}, and 4 for \emph{justlen}) that require an increasing number of transactions to reach an exploitable state plus another transaction to trigger a vulnerability (see \Cref{tab:bugbenchresults} for a summary).
Note that we cannot use real-world contracts here, as they do not allow us to scale the required number of transactions to trigger a bug.
Each variant of \emph{multi} and \emph{complex} contracts is parameterized given the number of transactions that are needed to trigger the bug.
For instance, contract \emph{multi}$_{10}$ requires 10 transactions (or sequential function calls) to reach an exploitable state.
The \emph{justlen} example is adapted from \citewauthor{Groce2021echidnaparade} and is parameterized over the length of an array that must be reached using operations such as \emph{push} and \emph{pop}.
We chose to insert a vulnerability in a function that simply triggers \emph{selfdestruct} of the contract when the exploitable state is reached.
This type of bug is widely supported by analysis tools and allows us to compare various tools according to the analysis time required to identify the bug.
The benchmarks are constructed to exercise the capability of solving constraints on the inputs (\emph{multi}, \emph{complex}) and the capability of moving a target into a certain internal state (\emph{justlen}).
Both capabilities are necessary for analyzing current smart contracts.
We provide more details on these benchmark contracts in \Cref{sec:benchdetails}.

We chose several state-of-the-art analysis tools that utilize different approaches to analysis, covering a large spectrum of analysis techniques (fuzzing, symbolic and concolic execution, and hybrids) and implementations.
More concretely, we compare \toolname with tEther~\cite{KruppR18teether}, MAIAN~\cite{NikolicKSSH18maian}, EthBMC~\cite{ FrankAH20ethbmc}, Manticore~\cite{MossbergMHGGFBD19manticore}, VeriSmart (SmarTest)~\cite{So2021verismart,So2021verismartsmartest}, Confuzzius~\cite{Torres2021confuzzius}, and Echidna~\cite{GriecoSCFG20echidna,Groce2021echidnaparade}.
We analyze our benchmark contracts with all these tools and measure the time until the bug is discovered with a global timeout of \SI{48}{\hour}.
We run all analysis tools within Docker containers on an Intel Xeon Gold 6230 CPU clocked at \SI{2.10}{\giga\hertz} with \SI{188}{\giga\byte} RAM.
We run the tools in parallel, keeping all physical CPU cores fully occupied, but do not utilize hyperthreaded cores.
All tools were executed on a single core, except those tools that support multi-core analysis, which we also run on 4 cores.
To run this experiment, we had to patch the tools tEther~\cite{KruppR18teether}, MAIAN~\cite{NikolicKSSH18maian}, and ConFuzzius~\cite{Torres2021confuzzius} such that they support longer transaction sequences.
We excluded EthBMC~\cite{FrankAH20ethbmc} from most of our experiments since we were not able to patch the bug that causes it to not report any vulnerabilities with transaction sequences longer than 3.
Moreover, we excluded ILF~\cite{HeBATV19ilf} because a machine learning-based fuzzer is unlikely to produce the \SI{256}{\bit} magic value constants used in the synthesized contracts.
We bound the number of transactions to consider by \num{32} to ensure that all tools execute at a reasonable pace while leaving enough room for failing/duplicated transactions.
For the Python-based tools, we utilize the PyPy JIT-compiler if we observe a speed-up during analysis (for MAIAN and teEther).
We perform 3 trials for all symbolic execution tools (where randomness does not play a big role) and at least 10 trials for all fuzzers.
In total, we spent approximately \SI{1300}{\day} of CPU time on this experiment.

Our results are summarized in \Cref{tab:bugbenchresults} and in the plots in \Cref{fig:otherplots}.
The plots show the log-scaled time required to solve the benchmark contracts with an increasing number of required transactions.
We omit those tools/results from the plots where all runs fail to identify a bug.
\toolname is the only analysis tool capable of solving all of the contracts in this benchmark dataset.
By adapting state-of-the-art fuzzing techniques, the performance of \toolname is comparable to symbolic execution when it comes to overcoming integer constraints.
\toolname performs better in the case of path-explosion-inducing code structures, such as loops or array handling.

Symbolic and concolic tools perform very well on the \emph{multi} benchmark because they explore the transaction ordering in a structured manner and can easily solve the integer-based path constraints using SMT solving.
However, we observe that most symbolic or concolic tools have trouble coping with the input-dependent loops in the \emph{complex} and \emph{justlen} benchmarks.

Classic fuzzing, as is performed by Echidna, fails to overcome the complex integer path constraints in \emph{multi} and \emph{complex}, but can generate long transaction sequences easily (e.g., on the \emph{justlen} benchmark).
Confuzzius and Smartian adopt a hybrid symbolic/fuzzing approach, where the analysis is driven by fuzzing, but inputs are also generated with symbolic constraint solving.
We believe that these tools perform poorly because they inherit the slow input generation from symbolic execution but perform probabilistic fuzzing of transaction sequences.
We also run a version of the \emph{multi$_{10}$} contract without any inputs/constraints.
It takes Confuzzius and Smartian, a mean of \num{2827} and \num{266} minutes, respectively, to find the correctly ordered transaction sequence.
In this experiment, it takes \toolname only about \num{2} minutes.

\subsection{Performance Ablation Study}%
\label{sec:eval:throughput}

\newcommand{\throughput}{\SI{}{\exec\per\second}}
\newcommand{\coverage}{cov \SI{}{\percent}}

\begin{table}[b]
	\centering
	\caption{Throughput measurements in average test case executions per second and mean code coverage. Best coverage is underlined.}
	\label{tab:throughput}
	\vspace{-0.5em}
	\resizebox{\linewidth}{!}{%
		\begin{tabular}{lr|rcrcrcrc}
			\toprule
			            &            & \multicolumn{2}{c}{Interp} & \multicolumn{2}{c}{AFL} & \multicolumn{2}{c}{EM} & \multicolumn{2}{c}{Full}                                                                               \\
			Contract    & LOC        & \throughput                & \coverage               & \throughput            & \coverage                & \throughput & \coverage              & \throughput & \coverage              \\
			\midrule

			Crowdsale   & \num{41}   & \num{13042}                & \num{79.5}              & \num{29688}            & \num{76.1}               & \num{19868} & \num{80.6}             & \num{25858} & \underline{\num{87.3}} \\
			multi$_{9}$ & \num{150}  & \num{12814}                & \num{43.7}              & \num{43053}            & \num{40.8}               & \num{20641} & \underline{\num{75.7}} & \num{25817} & \num{52.5}             \\
			IMBTC       & \num{664}  & \num{6880}                 & \num{36.6}              & \num{28372}            & \num{52.6}               & \num{18444} & \num{36.4}             & \num{34510} & \underline{\num{52.9}} \\
			PackSale    & \num{730}  & \num{7146}                 & \num{65.0}              & \num{32215}            & \num{67.5}               & \num{11952} & \num{60.2}             & \num{24672} & \underline{\num{75.2}} \\
			Spankchain  & \num{1048} & \num{6574}                 & \num{26.9}              & \num{19837}            & \num{47.0}               & \num{17245} & \underline{\num{50.5}} & \num{22698} & \num{41.7}             \\
			CryptoBets  & \num{1142} & \num{3174}                 & \num{30.8}              & \num{25122}            & \num{35.0}               & \num{10256} & \underline{\num{45.0}} & \num{12246} & \num{40.5}             \\

			\bottomrule
		\end{tabular}
	}
\end{table}

\let\throughput\undefined
\let\coverage\undefined

To evaluate the performance impact of various components of \toolname, such as our \emph{evm2cpp} compiler, we perform an ablation study concerning test case throughput and achieved code coverage.
We leverage a set of contracts consisting of a mix of real-world contracts and one of our benchmark contracts: the contracts Crowdsale~\cite{HeBATV19ilf} and the synthetic multi$_{10}$ represent simpler contracts, while the {IMBTC}~\cite{imbtc}, {SpankChain}~\cite{spankchain}, CryptoBets, and PackSale contracts are more complex real-world contracts.
We run each fuzzing configuration \num{20} times for \SI{10} minutes and record the average test case executions per second along with the EVM basic block coverage.
Our evaluation results are summarized in \Cref{tab:throughput}.
We provide further experiments with the \emph{multi}, \emph{complex}, and \emph{justlen} benchmark contracts in \Cref{sec:modebenchmarks}.

When we disable \emph{evm2cpp} and run the smart contract in the EVM interpreter provided by the \emph{eEVM} project (labeled \emph{Interp} in \Cref{tab:throughput}), we observe significantly lower test case throughput.
This directly translates to achieving less code coverage with the fuzzer, highlighting the importance of \emph{evm2cpp} in our design.
For the \emph{AFL} configuration, we disable the custom mutator and for the \emph{EM} configuration we disable AFL's mutations.
While the lightweight mutations performed by AFL++ result in the highest throughput, they are not aware of the input structure and often achieve worse code coverage.
We noticed that AFL++ alone fails to create combinations of transactions (see \Cref{sec:modebenchmarks}) because it lacks structural mutations.
In this case, basic block coverage is not a good metric since it does not account for different combinations of transactions.
The custom mutator alone sometimes cannot discover all interesting code paths due to worse throughput and the lack of AFL++'s advanced mutations.
Since fuzzing campaigns utilize multiple cores in practice, we take inspiration from \emph{ensemble fuzzing}~\cite{Chen2019enfuzz} and automatically launch different fuzzer configurations in parallel if multiple cores are available to \toolname (see \Cref{sec:impldetails}).

\paragraph{Comparison with other fuzzers}
We compare the test case throughput of other fuzzers with the throughput of \toolname.
Our measurement results show that the test case throughput of \toolname is larger by an order of magnitude when compared with other fuzzers.
Within our throughput benchmark set, \toolname has a mean throughput of \SI{24301}{\exec\per\second} ($\sigma = 7154$).
Echidna~\cite{GriecoSCFG20echidna} achieves a throughput of \num{189} ($\sigma = 183$) and a maximum throughput of \num{497} test cases per second.
Confuzzius~\cite{Torres2021confuzzius} has a mean throughput of \num{78} transactions per second ($\sigma = 25$).
Notice that the transaction throughput is always higher or equal to the test case throughput since test cases typically consist of multiple transactions.

\subsection{Code Coverage Comparison}%
\label{sec:eval:codecov}

\begin{table}[b]
	\center
	\caption{Number of targets, where \toolname statistically significantly outperforms ConFuzzius/ILF and vice versa on the top 100 targets in various complexity properties}
	\resizebox{\linewidth}{!}{%
		\begin{tabular}{@{}lccccc@{}}
			\toprule
			                       & \#LLOC  & \#funcs & \#comp  & \#branch \\
			\midrule
			\toolname~: ConFuzzius & 73 : 17 & 66 : 22 & 60 : 28 & 77 : 15  \\
			\toolname~: ILF        & 55 : 37 & 46 : 42 & 66 : 26 & 54 : 37  \\
			\bottomrule
		\end{tabular}
	}
	\label{tab:complex_contracts_top100}
\end{table}

We compare \toolname with the fuzzers ILF~\cite{HeBATV19ilf} and Confuzzius~\cite{Torres2021confuzzius} on a set of real-world smart contracts.
We cannot compare the fuzzers according to their time-to-bug, since
\begin{inparaenum}[(1)]
	\item there are no datasets available with both ground truth and realistically complex contracts, and
	\item the fuzzers' bug oracles differ too much to be comparable.
\end{inparaenum}
Instead, we focus on the fuzzers' capability of achieving code coverage, which is a necessary but not a sufficient condition to discover bugs.
We were not able to find a way to report basic block coverage in Echidna~\cite{GriecoSCFG20echidna}, which we exclude from the comparison.
We utilize a set of real-world smart contracts, extracted from the \emph{smartbugs-wild} dataset, which are supported by all fuzzers (see \Cref{sec:benchdetails} for more details).

We configure all fuzzers such that they behave reasonably similarly and offer comparable results.
Confuzzius achieves better code coverage when a smart contract contains hard-coded addresses.
Only Confuzzius generates transactions originating from those hard-coded addresses.
While this behavior does lead to good code coverage, the generated transaction sequences cannot actually be performed on the blockchain, as an arbitrary address cannot be impersonated.
For this reason, we disabled marking the \emph{caller}, i.e., the origin of a transaction, as an unconstrained symbolic value in Confuzzius.
Additionally, we had to patch Confuzzius's coverage reporting to take bytecode as input instead of source code.
We patched ILF to improve the reporting of code coverage and transaction inputs and replaced the existing threshold on the number of generated inputs with a time-based limit.
For \toolname, we enabled an over-approximating mode, where the data returned by all external calls is fuzzed and allow \toolname to generate calls originating from the contract creator.

To compare \toolname with the other fuzzers, we follow state-of-the-art recommendations for fuzzer evaluation~\cite{ARC2014STATS,KLE2018EVAL}.
We repeat every experiment 30 times to account for the randomness in the fuzzing process and limit the runtime to five minutes for each target.
We use the SENF~\cite{PA2021SENF} framework to calculate the required statistical evaluation metrics.
We find that \toolname performs better with statistical significance on 141 targets when compared with Confuzzius and 120 targets when we compare it with ILF.
In contrast, Confuzzius and ILF perform better on 83 and 112 of the target contracts, respectively (see \Cref{sec:evalappendix} for more details).
We conduct an additional evaluation on the top 100 most complex targets with respect to the number of logical lines of code, functions, comparisons, and branches.
As shown in \Cref{tab:complex_contracts_top100}, \toolname outperforms Confuzzius and ILF on the majority of contracts across all complexity properties.
Thus, we conclude that \toolname can handle increasingly complex Ethereum smart contracts better than existing fuzzers.

\subsection{Bug Detection Capabilities}%
\label{sec:eval:bugdetection}

To assess the bug detection capability of \toolname, we tested several real-world contracts obtained from prior studies~\cite{FrankAH20ethbmc,ZhouYXCY020eeg,bose2021sailfish,Cecchetti2021serif}.
Note that after initial testing, we concluded that existing benchmark datasets with ground truth for Solidity/Ethereum analysis tools~\cite{Durieux2020smartbugs, ghaleb2020soliditybuginjection} are not suitable for testing/comparing dynamic analysis tools such as fuzzers.
Existing benchmark datasets consist mostly of rather simple and very similar contracts.
For example, the curated reentrancy dataset of \citewauthor{Durieux2020smartbugs} features mostly honeypot contracts designed to be easily analyzed~\cite{Torres2019honeybadger}.
Similarly, the reentrancy bugs injected by \citewauthor{ghaleb2020soliditybuginjection} are too simplistic: Many of the injected bugs cannot be triggered by dynamic analysis tools (dead code) or are trivially exploitable (see \Cref{sec:datasetproblems}).
For this reason, we rely on prior studies on real-world contracts for evaluating the bug detection capabilities of \toolname.

\paragraph{Access Control Vulnerabilities}
Access control bugs such as an unprotected \emph{selfdestruct} have been widely investigated~\cite{KruppR18teether, NikolicKSSH18maian, FrankAH20ethbmc}.
We obtained a list of \num{2856} contracts that are vulnerable according to EthBMC~\cite{FrankAH20ethbmc}.
We export the blockchain state of the contracts at block number \num{9069000}, import the state into \toolname, and fuzz all contracts until the bug is discovered (with a timeout of \SI{20}{\minute}).
In total, \toolname detects a vulnerability in \num{2825} out of \num{2856} contracts.
On average it takes \toolname \SI{28.5}{\second} ($\sigma = 111.4$) until the bug is discovered.
For 18 contracts, \toolname detected no bug in our first run.
The remaining contracts had issues due to errors during state export.
Among the 18 contracts, we find that 5 are not vulnerable and have been mistakenly marked as vulnerable.
We run with a slightly earlier blockchain state and find that \toolname identifies another 4 vulnerable contracts.
The remaining 12 contracts contain bugs missed by \toolname caused by inefficiency when fuzzing without ABI information.

We also applied \toolname on \num{10356} contracts which EthBMC was unable to analyze because of timeouts after \SI{30}{\minute}, i.e., contracts that cannot be analyzed with bounded model checking.
In contrast, \toolname successfully processed all these contracts and achieves an average code coverage of \SI{72.4}{\percent} ($\sigma = 17.9$).
Furthermore, \toolname detected 85 vulnerable contracts in this set, of which we manually checked 18 contracts with verified source code and found 7 true vulnerabilities.
For the remaining 11 contracts, \toolname correctly identifies a transaction sequence to gain Ether.
However, these contracts intentionally implement features that allow any user to obtain Ether.
This is common for gambling contracts or contracts that pay out dividends (see \Cref{sec:discussion}).

\paragraph{Reentrancy Vulnerabilities}
We evaluated \toolname using a set of contracts vulnerable to reentrancy according to prior studies~\cite{sereum,ZhouYXCY020eeg,bose2021sailfish,Cecchetti2021serif}.
Furthermore, to give an intuition about detection capabilities we also provide results for Confuzzius~\cite{Torres2021confuzzius} and the static source code analyzer Slither~\cite{FeistGG19slither} in \Cref{tab:reentrancy}. %
In contrast to \toolname, which generates reentrancy exploits, Confuzzius and Slither feature a heuristic detection of reentrancy issues.
Slither defines any state update after an external call to be a potential reentrancy bug.
Similarly, Confuzzius defines a reentrancy bug as an external call, where some state variable is read before the call and written after the call.
Confuzzius does not actually generate transaction sequences that contain reentrant transactions.
We filter out wrongly detected reentrancy bugs by manually analyzing the reported contracts from prior studies~\cite{sereum,ZhouYXCY020eeg}.
Furthermore, many cases are trivial reentrancy bugs, summarized as \emph{Trivial-RE} in \Cref{tab:reentrancy}, which includes reentrancy honeypot contracts~\cite{Torres2019honeybadger} (see \Cref{sec:honeypotre}).

\begin{table}[b]
	\newcommand{\tta}{{\color{ForestGreen}{$\checkmark$}}}
	\newcommand{\tno}{{\color{red}{$\times$}}}
	\newcommand{\tfa}{{\color{orange}{$\thicksim$}}}
	\newcommand{\tmay}{\tfa}
	\newcommand{\tna}{N/A}
	\newcommand{\tun}{{\color{red}{?}}}
	\centering
	\vspace{-1em}
	\caption{Results for reentrancy issues for various analysis tools: False Alarms (\tfa), True Alarms (\tta), not applicable/incompatible (\tna), or as Missed Bug (\tno).}
	\label{tab:reentrancy}
	\resizebox{0.8\linewidth}{!}{%
		\begin{tabular}{l|ccc}
			\toprule
			Contract                                    & \toolname & Confuzzius & Slither \\
			\midrule
			Example \Cref{fig:hardre}                   & \tta      & \tfa       & \tfa    \\
			SpankChain~\cite{spankchain}                & \tta      & \tno       & \tta    \\
			DSEthToken~\cite{sereum}                    & \tta      & \tta       & \tna    \\
			TheDAO~\cite{sereum}                        & \tna      & \tta       & \tna    \\
			HODLWallet~\cite{hodlwallet,ZhouYXCY020eeg} & \tno      & \tta       & \tta    \\
			SysEscrow~\cite{sysescrow,ZhouYXCY020eeg}   & \tta      & \tno       & \tta    \\
			InstaDice~\cite{instadice,ZhouYXCY020eeg}   & \tta      & \tfa       & \tta    \\
			Trivial-RE~\cite{Torres2019honeybadger}     & \tta      & \tta       & \tta    \\
			\bottomrule
		\end{tabular}
	}

	\let\tfa\undefined
	\let\tta\undefined
	\let\tno\undefined
	\let\tna\undefined
	\let\tun\undefined
\end{table}

\toolname is highly effective in discovering all known reentrancy issues. 
However, the prototype implementation of \toolname does not yet support contract creation at runtime (see \Cref{sec:implementation:translation}). 
We noticed dynamic contract creation for ``the DAO'' contract, which is the reason \toolname does not detect the DAO reentrancy.
The \emph{HODLWallet} contract requires special attention: While this contract is vulnerable to a reentrancy bug, it cannot be exploited to gain Ether.
According to our analysis, this contract allows users to invest Ether into the contract, but the contract never returns all the invested Ether.
However, a reentrancy bug in the contract can be exploited to withdraw all previously invested Ether.
\toolname's bug oracle does not identify this as a reentrancy bug since no Ether can actually be gained.
For the \emph{InstaDice} contract, Confuzzius reports many additional reentrancy issues even when the contract calls into other trusted contracts instead.
\toolname, on the other hand, executes trusted contracts exported from the Ethereum node and produces a fully working exploit for the reentrancy bug without reporting false alarms.
Remarkably, \toolname is the \emph{only} dynamic analysis tool that is able to accurately identify real-world reentrancy issues such as the reentrancy bugs in the SpankChain and DSEthToken contracts.

We also evaluated against the reentrancy bugs recently discovered by \citewauthor{bose2021sailfish} with the \textsc{Sailfish} tool.
The study reports \num{26} contracts with true reentrancy bugs in the dataset.
However, we were able to confirm only \num{5} of these contracts to be vulnerable to Ether stealing with \toolname.
Our (manual) analysis on the remaining 21 contracts reveals that, while the contracts can be reentered in theory, all but one cannot be exploited (see \Cref{sec:sailfishre}).
We also identified one contract in this set that can be exploited because of an access control bug, not because of reentrancy.
This shows that accurate bug oracles, such as those used by \toolname, are less likely to produce false alarms and therefore give better feedback to smart contract developers by providing concrete transaction sequences that trigger the reentrancy bug.

\paragraph{Compositional Security}
We follow the evaluation of the Serif static analyzer~\cite{Cecchetti2021serif} to show the feasibility of detecting compositional security violations with \toolname.
We adapted the contracts from Serif's evaluation set such that they are deployable and exploitable in a realistic setting.
Where Serif relies on manual annotations to detect potential problems, we augment the contracts with assertions that are picked up by \toolname to detect vulnerabilities beyond Ether-stealing.
\toolname accurately generates a reentrancy attack for the Uniswap and Multi-DAO contracts in this dataset.
Furthermore, \toolname generates transactions that violate the assertions for the KV-Store and TownCrier contracts using reentrant transactions.
To evaluate \toolname's capabilities on a real-world example, we also fuzz the composition of Uniswap and IMBTC, which are exploitable due to a reentrancy bug~\cite{Torres21eyeofhorus}. 
We supply \toolname with the addresses of three contracts that should receive transactions (Uniswap, IMBTC and the ERC1820Registry) and export the state of these contracts at block number \num{9600000}, shortly before the first known attack.
We then fuzz this composition on \num{40} cores for a maximum of \SI{48}{\hour}.
We repeated the experiment \num{10} times and find that \toolname on average requires \SI{1}{\hour} and \SI{49}{\minute} ($\sigma \approx \SI{14}{\hour} \SI{35}{\minute}$, geomean $\SI{6}{\hour} \SI{55}{\minute}$) to generate a reentrancy exploit that
\begin{inparaenum}[(1)]
  \item registers an attacker contract in the \emph{ERC1820Registry} to allow \emph{ERC777} callbacks,
  \item buys \emph{IMBTC} tokens via the \emph{Uniswap} contract, and
  \item finally exploits the reentrancy to sell them again with a profit.
\end{inparaenum}

\section{Discussion / Related Work}%
\label{sec:discussion}
\vspace{-0.2em}
\paragraph{Fuzzing Structured Input}
On the protocol level, inputs for smart contracts are a byte string encoded according to the \emph{ABI} definition.
A fuzzer must provide inputs that are valid according to the ABI or the contract will stop execution early.
Symbolic execution tools handle this by utilizing the SMT solver to identify valid input data.
Most smart contract fuzzers~\cite{GriecoSCFG20echidna,NguyenP0L020sfuzz,HeBATV19ilf} use an approach akin to grammar-fuzzing~\cite{Burkhardt1967fromsyntax,Yang2011csmith} to generate inputs with the ABI.
While \toolname also utilizes the ABI to efficiently mutate transaction inputs, it does not solely rely on the \emph{ABI}.
\toolname leverages the lightweight mutations of its base fuzzer to mutate raw input.
Using the coverage feedback, \toolname discovers structurally valid inputs even without the ABI.
This allows \toolname to fuzz contracts where no source code or only an incomplete ABI is available.
For example, \toolname can fuzz inputs that contain byte strings that are further decoded elsewhere, e.g., because the data is forwarded to another smart contract.
\toolname also fuzzes the return data of external calls, which is ABI-encoded but not included in a contract's ABI.
However, fuzzing without ABI information is currently less well-optimized in \toolname (see \Cref{sec:modebenchmarks}).

\vspace{-0.2em}
\paragraph{Bug Oracles}
There is a wide spectrum of bug oracles in smart contracts.
Analysis tools define their own bug oracles, sometimes with slightly different definitions of the same bug classes.
Many analysis tools~\cite{Luu2016oyente,Torres2018osiris,Torres2021confuzzius,HeBATV19ilf,bose2021sailfish} feature bug oracles that indicate issues, but not necessarily a security vulnerability.
For example, detecting a data dependency on the block timestamp might be a sign of a bad attempt at using randomness, or it might be a legitimate use to implement a time-limited sale.
While such oracles also result in a larger number of alarms, they have the advantage that they can uncover a larger set of issues.
Developers can use these findings to improve code quality.
Other analysis tools implement Ether-based bug oracles for exploit generation~\cite{KruppR18teether,NikolicKSSH18maian,FrankAH20ethbmc} with few false alarms.
\toolname also utilizes such a bug oracle but also covers complex interactions (e.g., reentrancy).
The downside is that token-related vulnerabilities cannot be directly detected using this bug oracle.
However, similar to Echidna~\cite{GriecoSCFG20echidna}, \toolname supports using developer annotations as bug oracles, allowing \toolname to identify token-related bugs and other logic bugs.

\vspace{-0.2em}
\paragraph{Simulating Benign Interactions}
An important property of a fuzzer is whether it simulates the behavior of benign users, i.e., whether the fuzzer can generate transaction sequences of the form $(t_u, t_a, t_u', t_a', \ldots)$, where $t_u$ and $t_u'$ are from a benign user, and $t_a$ and $t_a'$ are from an attacker.
For example, many smart contracts implement the \emph{Owned} pattern: There is one Ethereum account that has special privileges for the contract.
If the fuzzer aims for optimal code coverage, then simulating arbitrary addresses is beneficial to reach code paths guarded by access control checks (e.g., implemented in Confuzzius~\cite{Torres2021confuzzius} and ILF~\cite{HeBATV19ilf}).
While this approach leads to good code coverage, it also entails more false alarms.
For example, in contracts with transferable ownership, the fuzzer will make the simulated owner transfer the ownership to an attacker account.
In turn, this would allow the attacker to drain the funds of the contract.
This is a false alarm since the real owner would never transfer ownership to an attacker.
However, this \emph{Owned} pattern is prevalent in smart contracts, making existing fuzzers that adopt this behavior produce false alarms.
We therefore opted to disable the simulation of any non-attacker-controlled accounts in \toolname.

\toolname only fuzzes return data that is attacker-controlled, i.e., data returned by callbacks.
By default, \toolname stops execution if an unknown contract is called.
\toolname requires all contract dependencies to be set up using a custom blockchain state.
Optionally, \toolname supports a mode that mutates data returned by any external call, similar to e.g., Confuzzius~\cite{Torres2021confuzzius}.
However, this mode is disabled by default to minimize false alarms.
For example, we observed that when testing a contract that relies on a token contract to manage user balances, mutating all returned data would result in impossible paths being executed, e.g., two subsequent calls to \emph{getBalance} returning different values.

\vspace{-0.2em}
\paragraph{False Alarms and Missed Bugs}
Like any fuzzer, \toolname provides neither sound nor complete analysis.
However, we designed \toolname as an exploit generator with few false alarms at the cost of missing some bugs.
Other analysis tools~\cite{bose2021sailfish,Torres2021confuzzius} focus on detecting reentrancy patterns, but do not generate an exploit.
As such, it is not clear whether their reported findings are true vulnerabilities, and often they are not (e.g., see our comparison with Sailfish in \Cref{sec:eval:bugdetection} and \Cref{sec:sailfishre}).
In contrast, \toolname only reports reentrancy attacks that are exploitable, leading to fewer false alarms compared to prior work.
The downside of \toolname's approach to reeentrancy is that it misses bugs that are not covered by the Ether-gains oracle or custom assertions.

The Ether-gains bug oracle does report false alarms in some edge cases.
In our evaluation, we identified several types of contracts that repeatedly lead to false alarms.
\begin{inparaenum}[(1)]
\item Gambling contracts, where \toolname identifies the right blockchain state and input such that the attacker always wins.
\item Token airdrops, where \toolname deterministically triggers the airdrop, which often results in Ether gains by selling the airdropped tokens again.
\item Interest pay-outs, where \toolname flags interest or dividends that are paid out over time.
\end{inparaenum}
In all these scenarios, there is a way to gain Ether from those contracts, which EF/CF correctly reports.
However, these are not considered vulnerabilities, as the contracts are intended to give out Ether.

We identified two major reasons for \toolname missing bugs:
\begin{inparaenum}[(1)]
  \item The bug cannot be uncovered by the Ether-gains bug oracle.
  \item The vulnerable state is not reachable within \toolname's EVM environment because some prerequisite is missing.
    For example, the target depends on a second contract not available to \toolname, or the contract requires additional state setup by the contract's owner (e.g., the \emph{Pausable} pattern).
\end{inparaenum}

\vspace{-0.2em}
\paragraph{Testing Multi-Contract Setups}
Smart contracts increasingly feature dependencies on third-party contracts.
Various recent incidents~\cite{Torres21eyeofhorus,creamfinance,revestfinance} show that multi-contract analysis is required for automatic analysis of complex DeFI applications that consist of compositions of smart contracts that have been independently developed.
Here, \emph{compositional security} is an important security property~\cite{Cecchetti2021serif}.
We show that \toolname can handle complex inputs and dependencies between transactions.
Similar to analysis tools~\cite{FrankAH20ethbmc,GriecoSCFG20echidna}, \toolname fully supports calling other contracts that act as dependencies.
Recently, Echidna~\cite{GriecoSCFG20echidna} introduced a \emph{multi-abi} mode, where the fuzzer is allowed to call functions on multiple smart contracts.
Similarly, \toolname supports generating transactions targeting multiple different smart contracts to identify issues due to unsuspected state changes in a contract's dependencies.
In contrast to Echidna, \toolname also supports reentrant transactions to different contracts, allowing it to detect compositional reentrancy such as the Uniswap/IMBTC issue.

\section{Conclusion}%
\label{sec:conclusion}

There is a demand for developing efficient and scalable techniques for security testing of smart contracts, which are being increasingly used to encode complex business logic on blockchain platforms.
We show that high-throughput fuzzing, as implemented in \toolname, improves on prior analysis tools, including increased code coverage, reduced time to discover bugs, the capability to model complex interactions (such as reentrancy), and the capability to analyze even complex contracts and compositions of contracts.
We show that several optimizations facilitate the high fuzzing throughput: translating contract bytecode to native code and employing efficient structural mutations on transaction sequences and the associated transaction inputs.
\toolname has comparable capabilities in solving complex input constraints as symbolic execution tools, while gracefully handling contracts that induce path explosion.
We release \toolname along with all of our benchmarks as open-source software.
We hope that this allows efficient automated testing of contracts and fosters additional research in smart contract security.

\section*{Acknowledgment}
Funded by the Deutsche Forschungsgemeinschaft (DFG, German Research Foundation) under Germany's Excellence Strategy - EXC 2092 CASA - 390781972, the German Federal Ministry of Education and Research (BMBF, project iBlockchain – 16KIS0901K), and the European Union (ERC, CONSEC, No. 101042266).
Views and opinions expressed are however those of the authors only and do not necessarily reflect those of the European Union or the European Research Council Executive Agency.
Neither the European Union nor the granting authority can be held responsible for them.
The authors would like to thank Alexander Meyring, Justin Nobles, and Simon Janzon for their contributions to the implementation of \toolname as part of their project group at University of Duisburg-Essen.

{%
	\printbibliography[heading=bibintoc]

@InProceedings{sereum,
  title        = "Sereum: Protecting Existing Smart Contracts Against
                 Re-Entrancy Attacks",
  booktitle    = "Proceedings of the Network and Distributed System
                 Security Symposium",
  series       = "{NDSS}",
  author       = "Michael Rodler and Wenting Li and Ghassan Karame and
                 Lucas Davi",
  year         = "2019",
  URL          = "https://www.ndss-symposium.org/ndss-paper/sereum-protecting-existing-smart-contracts-against-re-entrancy-attacks/",
}

@InProceedings{FrankAH20ethbmc,
  author       = "Joel Frank and Cornelius Aschermann and Thorsten
                 Holz",
  title        = "{ETHBMC:} {A} Bounded Model Checker for Smart
                 Contracts",
  booktitle    = "29th {USENIX} Security Symposium",
  publisher    = "{USENIX} Association",
  year         = "2020",
  URL          = "https://www.usenix.org/conference/usenixsecurity20/presentation/frank",
}

@InProceedings{KruppR18teether,
  author       = "Johannes Krupp and Christian Rossow",
  title        = "{teEther}: Gnawing at Ethereum to Automatically
                 Exploit Smart Contracts",
  booktitle    = "27th {USENIX} Security Symposium",
  publisher    = "{USENIX} Association",
  year         = "2018",
  URL          = "https://www.usenix.org/conference/usenixsecurity18/presentation/krupp",
}

@InProceedings{NikolicKSSH18maian,
  author       = "Ivica Nikolic and Aashish Kolluri and Ilya Sergey and
                 Prateek Saxena and Aquinas Hobor",
  title        = "Finding The Greedy, Prodigal, and Suicidal Contracts
                 at Scale",
  booktitle    = "Proceedings of the 34th Annual Computer Security
                 Applications Conference",
  series       = "{ACSAC}",
  ublisher     = "{ACM}",
  year         = "2018",
  DOI          = "10.1145/3274694.3274743",
}

@InProceedings{HeBATV19ilf,
  author       = "Jingxuan He and Mislav Balunovic and Nodar Ambroladze
                 and Petar Tsankov and Martin T. Vechev",
  title        = "Learning to Fuzz from Symbolic Execution with
                 Application to Smart Contracts",
  booktitle    = "Proceedings of the 2019 {ACM} {SIGSAC} Conference on
                 Computer and Communications Security",
  series       = "{CCS}",
  publisher    = "{ACM}",
  year         = "2019",
  DOI          = "10.1145/3319535.3363230",
}

@InProceedings{NguyenP0L020sfuzz,
  author       = "Tai D. Nguyen and Long H. Pham and Jun Sun and Yun Lin
                 and Quang Tran Minh",
  editor       = "Gregg Rothermel and Doo{-}Hwan Bae",
  title        = "{sFuzz}: an efficient adaptive fuzzer for solidity
                 smart contracts",
  booktitle    = "{ICSE} '20: 42nd International Conference on Software
                 Engineering",
  publisher    = "{ACM}",
  year         = "2020",
  DOI          = "10.1145/3377811.3380334",
}

@InProceedings{Jiang18contractfuzzer,
  author       = "Bo Jiang and Ye Liu and W. K. Chan",
  editor       = "Marianne Huchard and Christian K{\"{a}}stner and
                 Gordon Fraser",
  title        = "{ContractFuzzer}: fuzzing smart contracts for
                 vulnerability detection",
  booktitle    = "Proceedings of the 33rd {ACM/IEEE} International
                 Conference on Automated Software Engineering",
  series       = "{ASE}",
  publisher    = "{ACM}",
  year         = "2018",
  DOI          = "10.1145/3238147.3238177",
}

@InProceedings{Wustholz2020harvey,
  title        = "Harvey: a greybox fuzzer for smart contracts",
  booktitle    = "Proceedings of the 28th {ACM} Joint Meeting on
                 European Software Engineering Conference and Symposium
                 on the Foundations of Software Engineering",
  author       = "Valentin W{\"u}stholz and Maria Christakis",
  publisher    = "{ACM}",
  series       = "ESEC/FSE 2020",
  month        = nov,
  year         = "2020",
  DOI          = "10.1145/3368089.3417064",
}

@InProceedings{Tsankov2018securify,
  title        = "Securify: Practical security analysis of smart
                 contracts",
  author       = "Petar Tsankov and Andrei Marian Dan and Dana
                 Drachsler{-}Cohen and Arthur Gervais and Florian
                 B{\"{u}}nzli and Martin T. Vechev",
  booktitle    = "Proceedings of the 2018 {ACM} {SIGSAC} Conference on
                 Computer and Communications Security",
  series       = "{CCS}",
  year         = "2018",
  publisher    = "{ACM}",
}

@InProceedings{Luu2016oyente,
  title        = "Making Smart Contracts Smarter",
  booktitle    = "Proceedings of the 2016 {ACM} {SIGSAC} Conference on
                 Computer and Communications Security",
  author       = "Loi Luu and Duc-Hiep Chu and Hrishi Olickel and
                 Prateek Saxena and Aquinas Hobor",
  series       = "{CCS}",
  year         = "2016",
  publisher    = "{ACM}",
}

@InProceedings{KalraGDS18zeus,
  author       = "Sukrit Kalra and Seep Goel and Mohan Dhawan and Subodh
                 Sharma",
  title        = "{ZEUS:} Analyzing Safety of Smart Contracts",
  booktitle    = "25th Annual Network and Distributed System Security
                 Symposium",
  series       = "{NDSS}",
  publisher    = "The Internet Society",
  year         = "2018",
}

@Online{dao-attack,
  author       = "Christoph Jentzsch",
  title        = "{The History of the DAO and Lessons Learned}",
  URL          = "https://blog.slock.it/the-history-of-the-dao-and-lessons-learned-d06740f8cfa5",
  lastaccessed = "October 10, 2017",
}

@InProceedings{MossbergMHGGFBD19manticore,
  author       = "Mark Mossberg and Felipe Manzano and Eric Hennenfent
                 and Alex Groce and Gustavo Grieco and Josselin Feist
                 and Trent Brunson and Artem Dinaburg",
  title        = "Manticore: {A} User-Friendly Symbolic Execution
                 Framework for Binaries and Smart Contracts",
  booktitle    = "34th {IEEE/ACM} International Conference on Automated
                 Software Engineering",
  series       = "{ASE}",
  publisher    = "{IEEE}",
  year         = "2019",
  DOI          = "10.1109/ASE.2019.00133",
}

@InProceedings{SchneidewindGSM20ethor,
  author       = "Clara Schneidewind and Ilya Grishchenko and Markus
                 Scherer and Matteo Maffei",
  title        = "{eThor}: Practical and Provably Sound Static Analysis
                 of Ethereum Smart Contracts",
  booktitle    = "{ACM} {SIGSAC} Conference on Computer and
                 Communications Security",
  seris        = "{CCS}",
  publisher    = "{ACM}",
  year         = "2020",
  DOI          = "10.1145/3372297.3417250",
}

@InProceedings{GriecoSCFG20echidna,
  author       = "Gustavo Grieco and Will Song and Artur Cygan and
                 Josselin Feist and Alex Groce",
  title        = "Echidna: effective, usable, and fast fuzzing for smart
                 contracts",
  booktitle    = "29th {ACM} {SIGSOFT} International Symposium on
                 Software Testing and Analysis",
  series       = "{ISSTA}",
  publisher    = "{ACM}",
  year         = "2020",
  DOI          = "10.1145/3395363.3404366",
}

@InProceedings{Groce2021echidnaparade,
  title        = "echidna-parade: a tool for diverse multicore smart
                 contract fuzzing",
  booktitle    = "Proceedings of the 30th {ACM} {SIGSOFT} International
                 Symposium on Software Testing and Analysis",
  author       = "Alex Groce and Gustavo Grieco",
  series       = "{ISSTA}",
  year         = "2021",
  DOI          = "10.1145/3460319.3469076",
}

@InProceedings{Torres21eyeofhorus,
  author       = "Christof Ferreira Torres and Antonio Ken Iannillo and
                 Arthur Gervais and Radu State",
  title        = "The Eye of Horus: Spotting and Analyzing Attacks on
                 Ethereum Smart Contracts",
  booktitle    = "Financial Cryptography and Data Security - 25th
                 International Conference",
  series       = "{FC}",
  publisher    = "Springer",
  year         = "2021",
  DOI          = "10.1007/978-3-662-64322-8\_2",
}

@InProceedings{ZhouYXCY020eeg,
  author       = "Shunfan Zhou and Zhemin Yang and Jie Xiang and Yinzhi
                 Cao and Min Yang and Yuan Zhang",
  title        = "An Ever-evolving Game: Evaluation of Real-world
                 Attacks and Defenses in Ethereum Ecosystem",
  booktitle    = "29th {USENIX} Security Symposium",
  publisher    = "{USENIX} Association",
  year         = "2020",
}

@InProceedings{FeistGG19slither,
  author       = "Josselin Feist and Gustavo Grieco and Alex Groce",
  title        = "Slither: a static analysis framework for smart
                 contracts",
  booktitle    = "Proceedings of the 2nd International Workshop on
                 Emerging Trends in Software Engineering for
                 Blockchain",
  series       = "{WETSEB@ICSE}",
  publisher    = "{IEEE} / {ACM}",
  year         = "2019",
  DOI          = "10.1109/WETSEB.2019.00008",
}

@InProceedings{Fioraldi2020aflpp,
  title        = "{AFL++}: Combining incremental steps of fuzzing
                 research",
  booktitle    = "14th {USENIX} Workshop on Offensive Technologies",
  series       = "{WOOT}",
  author       = "Andrea Fioraldi and Dominik Maier and Heiko
                 Ei{\ss}feldt and Marc Heuse",
  year         = "2020",
}

@InProceedings{Torres2021confuzzius,
  author       = "Christof Ferreira Torres and Antonio Ken Iannillo and
                 Arthur Gervais and Radu State",
  title        = "{ConFuzzius}: {A} Data Dependency-Aware Hybrid Fuzzer
                 for Smart Contracts",
  booktitle    = "{IEEE} European Symposium on Security and Privacy",
  series       = "EuroS{\&}P",
  publisher    = "{IEEE}",
  year         = "2021",
  DOI          = "10.1109/EuroSP51992.2021.00018",
}

@Online{aflhomepage,
  title        = "{American Fuzzy Lop}",
  author       = "Michal Zalewski",
  URL          = "https://lcamtuf.coredump.cx/afl/",
  urldate      = "2022-04-28",
}

@InProceedings{Aschermann2019redqueen,
  title        = "{REDQUEEN}: Fuzzing with {Input-to-State}
                 Correspondence",
  booktitle    = "Proceedings 2019 Network and Distributed System
                 Security Symposium",
  series       = "{NDSS}",
  author       = "Cornelius Aschermann and Sergej Schumilo and Tim
                 Blazytko and Robert Gawlik and Thorsten Holz",
  publisher    = "Internet Society",
  year         = "2019",
  conference   = "Network and Distributed System Security Symposium",
  DOI          = "10.14722/ndss.2019.23371",
}

@InProceedings{Torres2018osiris,
  author       = "Christof Ferreira-Torres and Julian Sch{\"{u}}tte and
                 Radu State",
  title        = "Osiris: Hunting for Integer Bugs in Ethereum Smart
                 Contracts",
  booktitle    = "Proceedings of the 34th Annual Computer Security
                 Applications Conference",
  series       = "{ACSAC}",
  year         = "2018",
  DOI          = "10.1145/3274694.3274737",
}

@Online{solidity-checks-effects-interaction,
  title        = "Solidity: Security Considerations - Use the
                 Checks-Effects-Interactions Pattern",
  URL          = "https://docs.soliditylang.org/en/v0.8.7/security-considerations.html#use-the-checks-effects-interactions-pattern",
  urldate      = "2021-09-09",
}

@Online{solidty-reguard,
  title        = "Solidity by Example: Re-Entrancy",
  URL          = "https://solidity-by-example.org/hacks/re-entrancy/",
  urldate      = "2021-12-12",
}

@Online{il2cppunity,
  title        = "Unity Documentation: {IL2CPP} Overview",
  URL          = "https://docs.unity3d.com/Manual/IL2CPP.html",
  urldate      = "2021-09-14",
}

@Online{androidart,
  title        = "Android Runtime {(ART)} and Dalvik",
  URL          = "https://source.android.com/docs/core/runtime",
  urldate      = "2022-10-24",
}

@Article{Wimmer2019graalnative,
  author       = "Christian Wimmer and Codrut Stancu and Peter Hofer and
                 Vojin Jovanovic and Paul W{\"{o}}gerer and Peter B.
                 Kessler and Oleg Pliss and Thomas W{\"{u}}rthinger",
  title        = "Initialize once, start fast: application
                 initialization at build time",
  journal      = "Proc. {ACM} Program. Lang.",
  volume       = "3",
  number       = "{OOPSLA}",
  year         = "2019",
  DOI          = "10.1145/3360610",
}

@InProceedings{Scholz2016souffle,
  author       = "Bernhard Scholz and Herbert Jordan and Pavle Subotic
                 and Till Westmann",
  title        = "On fast large-scale program analysis in Datalog",
  booktitle    = "Proceedings of the 25th International Conference on
                 Compiler Construction",
  series       = "{CC}",
  year         = "2016",
  DOI          = "10.1145/2892208.2892226",
}

@InProceedings{Durieux2020smartbugs,
  title        = "Empirical review of automated analysis tools on 47,587
                 Ethereum smart contracts",
  booktitle    = "Proceedings of the {ACM/IEEE} 42nd International
                 Conference on Software Engineering",
  author       = "Thomas Durieux and Jo{\~a}o F Ferreira and Rui Abreu
                 and Pedro Cruz",
  series       = "ICSE",
  year         = "2020",
  DOI          = "10.1145/3377811.3380364",
}

@InProceedings{ghaleb2020soliditybuginjection,
  title        = "How Effective Are Smart Contract Analysis Tools?
                 Evaluating Smart Contract Static Analysis Tools Using
                 Bug Injection",
  author       = "Asem Ghaleb and Karthik Pattabiraman",
  booktitle    = "Proceedings of the 29th {ACM} {SIGSOFT} International
                 Symposium on Software Testing and Analysis",
  year         = "2020",
  series       = "{ISSTA}",
}

@InProceedings{Choi2021smartian,
  author       = "Jaeseung Choi and Doyeon Kim and Soomin Kim and
                 Gustavo Grieco and Alex Groce and Sang Kil Cha",
  title        = "{SMARTIAN:} Enhancing Smart Contract Fuzzing with
                 Static and Dynamic Data-Flow Analyses",
  booktitle    = "36th {IEEE/ACM} International Conference on Automated
                 Software Engineering",
  series       = "{ASE}",
  publisher    = "{IEEE}",
  year         = "2021",
  DOI          = "10.1109/ASE51524.2021.9678888",
}

@Misc{mythril,
  key          = "mythril",
  author       = "ConsenSys",
  title        = "Mythril v0.22.1",
  URL          = "https://github.com/ConsenSys/mythril",
}

@Misc{eEVM,
  key          = "eEVM",
  author       = "Microsoft",
  title        = "Enclave {EVM}",
  URL          = "https://github.com/microsoft/eEVM",
}

@Misc{libfuzzer,
  key          = "libfuzzer",
  author       = "{LLVM Compiler Infrastructure}",
  title        = "{libFuzzer -- a library for coverage-guided fuzz
                 testing}",
  URL          = "https://www.llvm.org/docs/LibFuzzer.html",
}

@InProceedings{Torres2019honeybadger,
  author       = "Christof Ferreira Torres and Mathis Steichen and Radu
                 State",
  title        = "The Art of The Scam: Demystifying Honeypots in
                 Ethereum Smart Contracts",
  booktitle    = "28th {USENIX} Security Symposium",
  year         = "2019",
  URL          = "https://www.usenix.org/conference/usenixsecurity19/presentation/ferreira",
}

@InProceedings{Xu2017fuzzprimitives,
  title        = "Designing New Operating Primitives to Improve Fuzzing
                 Performance",
  booktitle    = "Proceedings of the 2017 {ACM} {SIGSAC} Conference on
                 Computer and Communications Security",
  author       = "Wen Xu and Sanidhya Kashyap and Changwoo Min and
                 Taesoo Kim",
  series       = "{CCS}",
  year         = "2017",
  DOI          = "10.1145/3133956.3134046",
}

@InProceedings{Leijen2019mimalloc,
  author       = "Daan Leijen and Benjamin Zorn and Leonardo de Moura",
  title        = "Mimalloc: Free List Sharding in Action",
  booktitle    = "Programming Languages and Systems - 17th Asian
                 Symposium",
  series       = "{APLAS}",
  year         = "2019",
  DOI          = "10.1007/978-3-030-34175-6_13",
}

@InProceedings{Claessen2000quickcheck,
  author       = "Koen Claessen and John Hughes",
  title        = "{QuickCheck}: a lightweight tool for random testing of
                 Haskell programs",
  booktitle    = "Proceedings of the Fifth {ACM} {SIGPLAN} International
                 Conference on Functional Programming",
  series       = "{ICFP}",
  year         = "2000",
  DOI          = "10.1145/351240.351266",
}

@InProceedings{Xie2005symstra,
  author       = "Tao Xie and Darko Marinov and Wolfram Schulte and
                 David Notkin",
  title        = "Symstra: {A} Framework for Generating Object-Oriented
                 Unit Tests Using Symbolic Execution",
  booktitle    = "Tools and Algorithms for the Construction and Analysis
                 of Systems, 11th International Conference",
  series       = "{TACAS}",
  year         = "2005",
  DOI          = "10.1007/978-3-540-31980-1_24",
}

@InProceedings{Thummalapenta2011nz,
  author       = "Suresh Thummalapenta and Tao Xie and Nikolai Tillmann
                 and Jonathan de Halleux and Zhendong Su",
  title        = "Synthesizing method sequences for high-coverage
                 testing",
  booktitle    = "Proceedings of the 26th Annual {ACM} {SIGPLAN}
                 Conference on Object-Oriented Programming, Systems,
                 Languages, and Applications",
  series       = "{OOPSLA}",
  year         = "2011",
  DOI          = "10.1145/2048066.2048083",
}

@TechReport{Aitel2002spike,
  title        = "The advantages of block-based protocol analysis for
                 security testing",
  author       = "Dave Aitel",
  year         = "2002",
  URL          = "http://www.immunityinc.com/downloads/advantages_of_block_based_analysis.pdf",
}

@InCollection{Kaksonen2001protos,
  title        = "Software Security Assessment through Specification
                 Mutations and Fault Injection",
  booktitle    = "Communications and Multimedia Security Issues of the
                 New Century: {IFIP} {TC6} / {TC11} Fifth Joint Working
                 Conference on Communications and Multimedia Security
                 ({CMS'01})",
  author       = "Rauli Kaksonen and Marko Laakso and Ari Takanen",
  year         = "2001",
  DOI          = "10.1007/978-0-387-35413-2_16",
}

@InProceedings{PA2021SENF,
  author       = "David Paaßen and Sebastian Surminski and Michael
                 Rodler and Lucas Davi",
  title        = "My Fuzzer Beats Them All! Developing a Framework for
                 Fair Evaluation and Comparison of Fuzzers",
  booktitle    = "Proc. of European Symposium on Research in Computer
                 Security",
  series       = "{ESORICS}",
  publisher    = "Springer International Publishing",
  year         = "2021",
  DOI          = "10.1007/978-3-030-88418-5\_9",
}

@Article{ARC2014STATS,
  author       = "Andrea Arcuri and Lionel Briand",
  title        = "A Hitchhiker's guide to statistical tests for
                 assessing randomized algorithms in software
                 engineering",
  journal      = "Software Testing, Verification and Reliability",
  year         = "2014",
  DOI          = "10.1002/stvr.1486",
}

@InProceedings{KLE2018EVAL,
  author       = "George Klees and Andrew Ruef and Benji Cooper and
                 Shiyi Wei and Michael Hicks",
  title        = "Evaluating Fuzz Testing",
  booktitle    = "{ACM} Conference on Computer and Communications
                 Security {(CCS)}",
  year         = "2018",
  DOI          = "10.1145/3243734.3243804",
}

@Online{imbtc,
  URL          = "https://etherscan.io/address/0x3212b29E33587A00FB1C83346f5dBFA69A458923",
  title        = "etherscan.io: The Tokenized Bitcoin {(imBTC)}",
  urldate      = "2021-12-12",
}

@Online{spankchain,
  URL          = "https://etherscan.io/address/0xf91546835f756da0c10cfa0cda95b15577b84aa7",
  title        = "etherscan.io: {LedgerChannel} {(SpankChain)}",
  urldate      = "2021-12-12",
}

@Online{instadice,
  URL          = "https://etherscan.io/address/0xfe1b613f17f984e27239b0b2dccfb1778888dfae",
  title        = "etherscan.io: {InstaDice}",
  urldate      = "2021-12-12",
}

@Online{sysescrow,
  URL          = "https://etherscan.io/address/0x903643251af408a3c5269c836b9a2a4a1f04d1cf",
  title        = "etherscan.io: {SysEscrow}",
  urldate      = "2021-12-12",
}

@Online{hodlwallet,
  URL          = "https://etherscan.io/address/0x4a8d3a662e0fd6a8bd39ed0f91e4c1b729c81a38",
  title        = "etherscan.io: {HODLWallet}",
  urldate      = "2021-12-12",
}

@InProceedings{Chen2019enfuzz,
  title        = "{EnFuzz}: Ensemble Fuzzing with Seed Synchronization
                 among Diverse Fuzzers",
  booktitle    = "28th {USENIX} Security Symposium",
  author       = "Yuanliang Chen and Yu Jiang and Fuchen Ma and Jie
                 Liang and Mingzhe Wang and Chijin Zhou and Xun Jiao and
                 Zhuo Su",
  year         = "2019",
  URL          = "https://www.usenix.org/conference/usenixsecurity19/presentation/chen-yuanliang",
}

@InProceedings{bose2021sailfish,
  author       = "Priyanka Bose and Dipanjan Das and Yanju Chen and Yu
                 Feng and Christopher Kruegel and Giovanni Vigna",
  title        = "{SAILFISH:} Vetting Smart Contract State-Inconsistency
                 Bugs in Seconds",
  booktitle    = "43rd {IEEE} Symposium on Security and Privacy",
  series       = "S\&P",
  publisher    = "{IEEE}",
  year         = "2022",
  DOI          = "10.1109/SP46214.2022.9833721",
}

@Online{phashmap,
  author       = "Gregory Popovitch",
  title        = "The Parallel Hashmap",
  year         = "2019",
  day          = "10",
  month        = mar,
  URL          = "https://greg7mdp.github.io/parallel-hashmap/",
  urldate      = "2022-04-22",
}

@Article{Burkhardt1967fromsyntax,
  title        = "Generating test programs from syntax",
  author       = "W H Burkhardt",
  month        = mar,
  year         = "1967",
  DOI          = "10.1007/BF02235512",
}

@InProceedings{Yang2011csmith,
  title        = "Finding and understanding bugs in {C} compilers",
  booktitle    = "Proceedings of the 32nd {ACM} {SIGPLAN} Conference on
                 Programming Language Design and Implementation
                 ({PLDI})",
  author       = "Xuejun Yang and Yang Chen and Eric Eide and John
                 Regehr",
  publisher    = "Association for Computing Machinery",
  month        = jun,
  year         = "2011",
  DOI          = "10.1145/1993498.1993532",
}

@InProceedings{Schneidewind2020goodbadugly,
  title        = "The Good, The Bad and The Ugly: Pitfalls and Best
                 Practices in Automated Sound Static Analysis of
                 Ethereum Smart Contracts",
  booktitle    = "Leveraging Applications of Formal Methods,
                 Verification and Validation: Applications",
  author       = "Clara Schneidewind and Markus Scherer and Matteo
                 Maffei",
  publisher    = "Springer International Publishing",
  year         = "2020",
  DOI          = "10.1007/978-3-030-61467-6_14",
}

@InProceedings{So2021verismartsmartest,
  author       = "Sunbeom So and Seongjoon Hong and Hakjoo Oh",
  editor       = "Michael Bailey and Rachel Greenstadt",
  title        = "{SmarTest}: Effectively Hunting Vulnerable Transaction
                 Sequences in Smart Contracts through Language
                 Model-Guided Symbolic Execution",
  booktitle    = "30th {USENIX} Security Symposium",
  publisher    = "{USENIX} Association",
  year         = "2021",
  URL          = "https://www.usenix.org/conference/usenixsecurity21/presentation/so",
}

@InProceedings{So2021verismart,
  author       = "Sunbeom So and Myungho Lee and Jisu Park and Heejo Lee
                 and Hakjoo Oh",
  title        = "{VERISMART:} {A} Highly Precise Safety Verifier for
                 Ethereum Smart Contracts",
  booktitle    = "2020 {IEEE} Symposium on Security and Privacy",
  series       = "SP",
  publisher    = "{IEEE}",
  year         = "2020",
  DOI          = "10.1109/SP40000.2020.00032",
}

@Online{creamfinance,
  URL          = "https://medium.com/cream-finance/c-r-e-a-m-finance-post-mortem-amp-exploit-6ceb20a630c5",
  urldate      = "2022-06-07",
  title        = "{C.R.E.A.M}. Finance Post Mortem: {AMP} Exploit",
  author       = "C. R. E. A. M Finance",
  year         = "2021",
  month        = sep,
  day          = "1",
}

@Online{revestfinance,
  URL          = "https://blocksecteam.medium.com/revest-finance-vulnerabilities-more-than-re-entrancy-1609957b742f",
  title        = "Revest Finance Vulnerabilities: More than
                 Re-entrancy",
  author       = "BlockSec",
  year         = "2022",
  day          = "31",
  month        = mar,
  urldate      = "2022-06-07",
}

@InProceedings{Cecchetti2021serif,
  author       = "Ethan Cecchetti and Siqiu Yao and Haobin Ni and Andrew
                 C. Myers",
  title        = "Compositional Security for Reentrant Applications",
  booktitle    = "42nd {IEEE} Symposium on Security and Privacy",
  series       = "S\&P",
  publisher    = "{IEEE}",
  year         = "2021",
  DOI          = "10.1109/SP40001.2021.00084",
}

@Article{Manes2021fuzzingsurvey,
  title        = "The art, science, and engineering of fuzzing: A
                 survey",
  author       = "Valentin J M Manes and Hyungseok Han and Choongwoo Han
                 and Sang Kil Cha and Manuel Egele and Edward J Schwartz
                 and Maverick Woo",
  journal      = "IEEE Trans. Software Eng.",
  publisher    = "Institute of Electrical and Electronics Engineers
                 (IEEE)",
  volume       = "47",
  number       = "11",
  pages        = "2312--2331",
  month        = nov,
  year         = "2021",
  ISSN         = "0098-5589, 1939-3520",
  DOI          = "10.1109/tse.2019.2946563",
}

@Online{parity-postmorterm,
  title        = "A Postmortem on the Parity Multi-Sig Library
                 Self-Destruct",
  author       = "Parity Technologies",
  year         = "2017",
  day          = "15",
  month        = nov,
  URL          = "http://paritytech.io/a-postmortem-on-the-parity-multi-sig-library-self-destruct",
  urldate      = "2020-06-03",
}
}

\appendices
\crefalias{section}{appendix}

\section{Implementation Details}%
\label{sec:impldetails}

\paragraph{EVM Changes}
Similar to many dynamic analysis tools we use a custom Ethereum virtual machine (EVM) environment.
This allows us to optimize the EVM for fuzzing by adding instrumentation and bug oracles.
We used the open-source \emph{eEVM} project~\cite{eEVM} as the base for our EVM environment.
We added an implementation of several newer EVM opcodes, missing features, and various minor fixes.
Furthermore, we replaced the usage of \Cpp exceptions with return values in hot code paths.
This results in considerably better performance since fuzzing tends to frequently exercise the error handling paths of the smart contract code.
We also switched to a more optimized hashmap implementation~\cite{phashmap} and use \emph{mimalloc}~\cite{Leijen2019mimalloc} as the default allocator.

\paragraph{Test Case Format}
\Cref{fig:hardeattack} depicts an example of the test case format used by \toolname.
During fuzzing \toolname utilizes a custom binary format that can be mutated by the base fuzzer using normal bit-flipping mutations without inducing parsing failures.
However, for introspection we convert the test cases to a yaml-based textual representation, which is shown in \Cref{fig:hardeattack}.
Here the transactions are depicted as a sequence of transactions.
However, the second transaction specifies a return header with the \emph{reenter} flag set to two.
This means that the following two transactions are executed as reentrant calls.

In \toolname, every transaction has up to 255 associated \emph{return-headers}.
Each return header specifies what the simulated attacker contracts do when they are called by the target contract, i.e., a \emph{callback}.
The return header contains the return value, the return data, and the reenter field.
Mutating the reenter field allows the fuzzer to explore different shapes of call trees that represent different kinds of reentrancy attacks.
\Cref{fig:remut} depicts several examples of call-trees that \toolname is capable of generating.

\paragraph{AFL++ Integration}
We patched AFL++ for optimal integration with our custom mutator.
Our patches make AFL++ report an internal performance score to our custom mutator.
The custom mutator can then select the number of fuzzing rounds for a given test case based on this score.
Depending on the size and complexity of the test case, we apply different types of mutations and a different number of fuzzing rounds in the custom mutator.

We ensure that AFL++ extensively uses the structured trimming provided by our custom mutator.
We found that structured trimming is beneficial to the fuzzing process in \toolname.
Additionally, we utilize the trimming step of AFL++ to update the internal test case queue of our custom mutator.
This allows us to only add trimmed test cases to the internal queue.
In contrast to AFL++'s queue, we keep the complete internal queue in memory and use it for efficient structural splicing operations.

We also extend AFL++ with an additional manual feedback API with a function that allows to reserve a larger part of AFL++'s coverage map for direct feedback.
This is used by our modifications to the \emph{eEVM} runtime to provide explicit feedback on two properties of the execution.
First, we provide explicit feedback on the progress of solving comparison operators.
For example, for the \texttt{EQ} opcode we provide explicit coverage feedback to AFL++ for every \SI{64}{\bit} part of the \SI{256}{\bit} EVM-native integer that is equal.
This allows AFL++'s \emph{cmplog} mode, which implements input-to-state correspondence on the \SI{64}{\bit} comparison level, to effectively solve fuzzing roadblocks on EVM-native \SI{256}{\bit} level.
Similarly, we provide explicit coverage feedback to AFL++ whenever a contract executes in a reentrant call.
This allows the fuzzer to distinguish a reentrant execution from a normal execution, i.e., a simple form of context-sensitive coverage.
We found this beneficial for AFL++ to keep both reentrant and non-reentrant variants of the same transaction sequence in the queue.

When launching AFL++, we disable the byte-level \emph{auto-dict} feature since it is superseded by our replacement acting on the EVM bytecode level.
Re-implementing the \emph{auto-dict} feature in \emph{evm2cpp} results in significantly smaller and more useful dictionaries than relying on AFL++'s auto-dict mode, which scans the final binary for constants and also picks up irrelevant data, such as strings only relevant for the \emph{eEVM} runtime.

\paragraph{Multi-Core Fuzzing}
While AFL++ has a single-threaded design, it is capable of synching with other instances of AFL++ (and even other fuzzers) via the filesystem.
\toolname inherits the same technique, and the wrapper scripts we provide as part of \toolname can automatically launch multiple AFL++ instances.
Generally, it is recommended to launch AFL++ using multiple different configurations when using multiple cores~\cite{Fioraldi2020aflpp}.
We adapt these recommendations to \toolname.
When running on 4 or more cores, we launch the following configurations:
\begin{enumerate}
	\item A main instance with AFL++'s deterministic mutation stages enabled.
	\item A compare solver instance, with input-to-state~\cite{Aschermann2019redqueen} and EVM-level compare tracing enabled.
	\item One instance fuzzes only with the custom mutator.
	\item The remaining cores utilize AFL++'s lightweight havoc mutations and our custom mutator.
\end{enumerate}

\paragraph{Other Bug Oracles}
The standard bug oracles are implemented inside of the \emph{eEVM} runtime code.
To implement a new bug oracle, one has to modify the \Cpp implementation of the runtime code.
For performance reasons, \toolname does not rely on heavyweight program analysis techniques such as taint tracking to implement bug oracles.
Instead, the bug oracles in \toolname are limited to detecting bugs based on the state of the simulated Ethereum blockchain.
However, we believe that the existing bug oracles supported by \toolname already cover a large set of use cases.
Furthermore, developers can use custom properties or the event mechanism to implement custom bug oracles directly in Solidity code.

Currently, \toolname supports optional fuzzing modes, which are also used by industry fuzzers such as Echidna~\cite{GriecoSCFG20echidna} or Mythril~\cite{mythril,Wustholz2020harvey}.
For example, \toolname supports property-based fuzzing with an interface that is fully compatible with the Echidna fuzzer.
The developer specifies a property of the contract that must always hold, i.e., an invariant of the contract code.
Such properties are specified as a Solidity function that returns whether the property is currently true or false.
After every executed transaction, \toolname calls the configured property functions and checks whether the return value signals a violated property.
Similarly, \toolname can utilize the EVM event logging and error propagation mechanisms to detect bugs.
The smart contract developer emits a certain event whenever a bug is triggered.
Whenever this event is logged during fuzzing, \toolname will consider the execution to trigger a bug and report it.
Similarly, Solidity version $0.8$ or above report special error messages to the caller, whenever an assertion is violated or an integer overflow happens.
If configured, \toolname picks up these special error message return codes as a bug and reports it.
This way \toolname can be utilized to fuzz for more than Ether-based bugs and also uncover contract-specific logic bugs.
\Cref{fig:propexample} shows an example for an invariant specified as a solidity function.
\toolname checks whether this function returns true after every executed transaction.

\begin{figure}[t]
  \centering
  \inputminted{solidity}{figures/property_example.sol}
  \caption{\toolname supports property-based fuzzing. The developer specifies a custom function that checks a property of a smart contract that must always hold. \toolname repeatedly calls this function after every executed transaction to check whether the property still holds and reports a bug if not.}%
  \label{fig:propexample}
\end{figure}

\paragraph{Comparison of Bug Oracles}
Previous analysis tools often implement a wide variety of bug oracles~\cite{Luu2016oyente, NikolicKSSH18maian, Torres2021confuzzius, Choi2021smartian} to detect security vulnerabilities, code smells, and other potentially interesting properties of the code.
However, the definition of the bug oracles and what oracles should be considered as a security vulnerability differ across the literature.
We identify the \emph{unprotected selfdestruct} bug oracles as one of the few oracles that are recognized in almost all analysis tools.
We utilize this bug oracle in our benchmarks (see \Cref{sec:eval} and \Cref{sec:benchdetails}).
In \toolname, we focus on Ether gains as our primary bug oracle, as it features the least number of false alarms in practice.
However, this single bug oracle in \toolname actually maps to multiple bug oracles in other tools.
Furthermore, we implement several additional optional bug oracles that can be used in \toolname.
We show a comparison of supported bug oracles in \Cref{tab:oraclecompare}.
In the following, we discuss some of the bug oracles in more detail.

\begin{table}[b]
	\newcommand{\tno}{{\color{red}{$\times$}}\xspace}
	\newcommand{\tyes}{{\color{ForestGreen}{$\checkmark$}}\xspace}
	\newcommand{\tpossible}{{\color{ForestGreen}{$\checkmark^{\ast}$}}\xspace}
	\newcommand{\tnewsol}{{\color{orange}{$\checkmark^{\dagger}$}}\xspace}
	\newcommand{\tindirect}{{\color{orange}{$\times^{\ddag}$}}\xspace}
	\centering
	\caption{Comparison of bug oracles in various fuzzing-based analysis tools with the bug oracles available in \toolname.
		\tyes fully supported.
		\tno not supported.
		\tpossible supported but not enabled by default.
		\tnewsol only supported for contract compiled with Solidity version $> 0.8$.
		\tindirect only if it leads to triggering another bug oracle.
	}%
	\label{tab:oraclecompare}
	\resizebox{\linewidth}{!}{%
		\begin{tabular}{lcccc}
			\toprule
			Bug Name                            & \toolname             & Confuzzius~\cite{Torres2021confuzzius} & Smartian~\cite{Choi2021smartian} & Echidna~\cite{ GriecoSCFG20echidna} \\
			\midrule
			Assertion Failure                   & \tnewsol              & \tyes                                  & \tyes                            & \tpossible                          \\
			Arbitrary Write                     & \tindirect            & \tno                                   & \tyes                            & \tindirect                          \\
			Block State Dependency              & \tindirect            & \tyes                                  & \tyes                            & \tindirect                          \\
			Control-flow Hijack (\texttt{JUMP}) & \tindirect            & \tno                                   & \tyes                            & \tindirect                          \\
			Custom Event Oracle                 & \tpossible            & \tno                                   & \tno                             & \tpossible                          \\
			Custom Property Checking            & \tpossible            & \tno                                   & \tno                             & \tpossible                          \\
			Ether Gains                         & \tyes                 & \tno                                   & \tno                             & \tno                                \\
			Integer Overflow                    & \tindirect / \tnewsol & \tyes                                  & \tyes                            & \tindirect                          \\
			Leaking Ether                       & \tpossible            & \tyes                                  & \tyes                            & \tno                                \\
			Locking Ether                       & \tno                  & \tyes                                  & \tyes                            & \tno                                \\
			Multiple Send                       & \tno                  & \tno                                   & \tyes                            & \tno                                \\
			Reentrancy                          & \tindirect            & \tyes                                  & \tyes                            & \tno                                \\
			Require Violation                   & \tno                  & \tno                                   & \tpossible                       & \tno                                \\
			Transaction Origin Use              & \tindirect            & \tno                                   & \tyes                            & \tindirect                          \\
			Transaction Order Dependency        & \tno                  & \tyes                                  & \tno                             & \tno                                \\
			Unsafe Delegatecall                 & \tyes                 & \tyes                                  & \tyes                            & \tno                                \\
			Unprotected Selfdestruct            & \tpossible            & \tyes                                  & \tyes                            & \tpossible                          \\
			Un/Mishandled Exception             & \tindirect            & \tyes                                  & \tyes                            & \tno                                \\
			\bottomrule
		\end{tabular}
	}
	\let\tyes\undefined
	\let\tno\undefined
	\let\tnewsol\undefined
	\let\tindirect\undefined
	\let\tpossible\undefined
\end{table}

\emph{Locking Ether} is fundamentally a liveness property.
In general, liveness properties are hard to prove with a fuzzer.
A standard fuzzing approach can only show that a certain code path \emph{can} be reached.
However, to accurately report locked Ether, the fuzzer would have to show that a certain code path cannot be reached.
For this reason, many fuzzers actually implement a static analysis approach to detecting locked Ether.
For example, ILF~\cite{HeBATV19ilf} and Confuzzius~\cite{Torres2021confuzzius} simply scan the contract for any instruction that can, in theory, send Ether.
However, they do not verify that the instruction can actually be executed.
For this reason, this approach will only detect simple cases of locked Ether.

\emph{Leaking Ether} and \emph{Ether Gains} are two very related bug oracles.
Both attempt to identify bugs where the contract can be used to send Ether to some unrelated address.
\toolname supports detecting leaking Ether but disables the bug oracle by default.
The idea is that the attacker can trick the contract into sending Ether to some contract that has no previous relationship with the contract.
However, many contracts support transferring Ether indirectly to another address as a feature.
For example, all token contracts must support transferring tokens to arbitrary addresses as a feature.
In contrast, \toolname uses Ether gains as a bug oracle that covers more realistic cases.
\toolname will report an Ether gain bug whenever the sum of the Ether balances of all attacker-controlled accounts exceeds the initial sum of balances.
This allows \toolname a wider and more realistic set of issues.
For example, the vulnerability depicted in \Cref{fig:leoracleproblem} is detected neither by Confuzzius~\cite{Torres2021confuzzius} nor by Smartian~\cite{Choi2021smartian}.
However, since \toolname simulates multiple attacker-controlled accounts, it will quickly generate a transaction originating from the first account and leaking the Ether to a second attacker-controlled address.

Most analysis tools feature explicit \emph{Reentrancy} bug oracles.
In contrast, \toolname does not feature an explicit detector for reentrancy but simply generates reentrant transaction sequences that trigger other bug oracles, such as Ether gains.
In this way, \toolname detects reentrancy bugs, but only if they are actually exploitable.
In contrast to other analysis tools, this leads to fewer false alarms, e.g., when encountering manual reentrancy locking~\cite{sereum}.

By default, \toolname reports \emph{unprotected selfdestruct} only if the selfdestruct will transfer the remaining Ether of the target contract to the attacker, i.e., the address parameter of the selfdestruct is controlled by the attacker.
Optionally, \toolname can also report \emph{DoS}-style unprotected selfdestructs, i.e., if the selfdestruct can be triggered by anyone, but always targets a trusted address such as the owner.
This style of detection is featured in most other analysis tools.

With Solidity versions $> 0.8$, contracts feature automatic integer overflow checking and proper assertion violation reporting.
\toolname supports this new Solidity exception mechanism to signal errors to the fuzzer.
Previously, \emph{assert} statements were implemented with the \emph{INVALID} opcode that also triggers a transaction revert.
However, earlier contracts (pre $0.4$) also used this to implement failure of input sanitization, which makes it hard to reliably distinguish between a regularly failing transaction and a assertion violation across Solidity versions.
We expect developers to use newer Solidity versions for newly deployed contracts.
We opted to support only the new Solidity exception mechanism, which allows us to reliably detect internal errors in a smart contract.
This includes memory allocation failure, integer overflows, and internal assertions.
Developers can utilize \toolname for general robustness testing of their newly deployed smart contracts.

\paragraph{En/Decoding the ABI}
We use the \emph{ethabi} Rust library to parse the JSON-based ABI definition, which allows us to en- and decode the ABI format expected by the smart contract.
Sometimes the base fuzzer will break the ABI encoding, which results in the custom mutator attempting to decode an extremely large input data.
In fact, during development of \toolname, we uncovered and fixed two bugs in the \emph{ethabi} library such that it would not result in an irrecoverable error when attempting to decode broken ABI-encoded data.
However, we still set an \SI{8}{\kilo\byte} limit to the number of bytes we attempt to decode.
This guards against spending too much time on attempting to decode an unusually large input byte-string of a transaction.
Since it is unlikely a valid or useful input for the smart contract under test, it is preferable to avoid the lengthy decoding process altogether.
In these pathological cases, we fall back on the random mutations provided by the base fuzzer.

\paragraph{Additional Tooling}
Based on our custom mutator code, we additionally implemented several tools that proved to be useful for smart contract fuzzing.
For instance, \toolname also features a test case minimizer that performs structural minimization on a test case, allowing an analyst to reduce the size of a test case.
Furthermore, \toolname integrates a translator between our binary test case format and a human-readable \emph{yaml}-based format, allowing for easy manual modification of test cases.
\toolname can also convert a test case into a Solidity attack contract that can be deployed within a blockchain environment to study the generated attack.
These tools help an analyst performing root cause analysis given a test case that triggers a bug.

\section{Problems with Existing Datasets}%
\label{sec:datasetproblems}

By now, there are multiple attempts to create standardized datasets to evaluate smart contract analysis tools~\cite{Durieux2020smartbugs,ghaleb2020soliditybuginjection}.
However, they lack diversity of bugs, especially with respect to reentrancy bugs.
They often only contain the simple same-function reentrancy pattern and contain many honeypot contracts (see \Cref{sec:honeypotre}).

\citewauthor{ghaleb2020soliditybuginjection} attempt to synthesize a benchmark dataset using artificial bug injection.
Unfortunately, we find that these synthesized datasets are not suitable to test reentrancy detection based on fuzzing or symbolic execution.
The injected reentrancy bugs focus on same-function reentrancy and do not cover more complex reentrancy patterns, such as cross-function reentrancy~\cite{sereum}.
Furthermore, the injected patterns themselves are suboptimal.
For example, many of the injected reentrancy bugs do not \emph{require} a reentrant call to be exploitable.
\Cref{fig:solidifi1} shows an injected bug that is prone to reentrancy.
However, the function can be identified as vulnerable without resorting to a reentrancy attack.
For example, MAIAN~\cite{NikolicKSSH18maian} detects an Ether leaking vulnerability in this function because the injected function will unconditionally send Ether (line 4) the first time a caller calls the function.
Reentrancy is only necessary to repeatedly leak Ether.
A second type of injected reentrancy bug is shown in \Cref{fig:solidifi2}.
This reentrancy bug can never be triggered during fuzzing or symbolic execution.
The injected vulnerable function contains a guard, the \emph{require} statement in line 6, that cannot be satisfied, because it accesses a storage variable that is never modified in the contract's code.
This makes the reentrant call practically dead code.
We therefore conclude that the reentrancy bugs injected as part of the \emph{SolidiFI} project are not suitable to test dynamic analysis tools, i.e., fuzzers or symbolic execution-based analyzers.

\begin{figure}[t]
	\inputminted{solidity}{./figures/solidifi_re_1.sol}
	\caption{An injected reentrancy bug, which is also detected by bug oracles that only identify leaking Ether without considering reentrancy.}
	\label{fig:solidifi1}
  \vspace{-1em}
\end{figure}

\begin{figure}[t]
	\inputminted{solidity}{./figures/solidifi_re_2.sol}
	\caption{An injected reentrancy bug that cannot be triggered because of a guarding condition that cannot be satisfied.}
	\label{fig:solidifi2}
  \vspace{-1em}
\end{figure}

\section{Details on Benchmarks}%
\label{sec:benchdetails}

\paragraph{Scalability Benchmark Dataset}
We create a set of contracts that can be used to evaluate tools with respect to their scalability to generating long transaction sequences.
Here we utilize a time-to-bug approach to benchmarking.
We therefore require contracts with ground truth and bugs that are widely supported by analysis tools.
We opted to create our own set of benchmark contracts.
Furthermore, we devise multiple variants of the same contracts, allowing us to measure scalability to longer transaction sequences in a more fine-grained manner.

To this end, we automatically synthesize the \emph{multi} contracts as follows: For each function, several equality or i-equality constraints are enforced on up to six integer arguments before setting a boolean internal state variable.
These benchmark contracts test the capability of solving input constraints across multiple transactions.
The functions of the contract must be called in the right order and with the right inputs to trigger a \emph{selfdestruct} of the contract.
The synthesized contracts favor symbolic execution tools because there is no potential for path explosion within the functions.
This is because they do not contain any control-flow statements except for error handling.

The three \emph{complex} contracts consist of a manual adaptation of the \emph{multi} contracts with more varying constraints on the input and the internal state of the smart contract (e.g., requiring an array input of a certain length or requiring the \emph{sha3} hash of a fixed value as input).
The idea is to add more complex and diverse input requirements that are more akin to real-world smart contracts.
We only use three different variants of these contracts because they are created manually.

\paragraph{Code Coverage Dataset}
For our code coverage experiments in \Cref{sec:eval:codecov}, we utilize a set of real world smart contracts.
To create a realistic set of smart contracts, we first analyzed the \emph{smartbugs-wild} dataset~\cite{Durieux2020smartbugs} and ranked the contracts according to their peak Ether balance (as reported by \url{etherscan.io}).
Ranking according to peak Ether balance naturally excludes toy or test contracts and creates a dataset that is focused on contracts that are actually in use.

We then processed the top \num{1000} contracts according to this ranking and created a minset of those contracts that are supported by all fuzzers.
For example, the contract must not require constructor parameters such that the contracts can easily be deployed in every fuzzer.
The resulting dataset consists of a set of 253 contracts from the top 1000 contracts of the \emph{smartbugs-wild} dataset that offer a certain level of code diversity.

\section{Ablation Study: Scalability}%
\label{sec:modebenchmarks}

\begin{figure*}[ht]
	\centering
	\begin{subfigure}[b]{0.4\linewidth}  %
		\graphicspath{{figures/}}
		\def\svgwidth{\linewidth}
		\begin{scriptsize}
			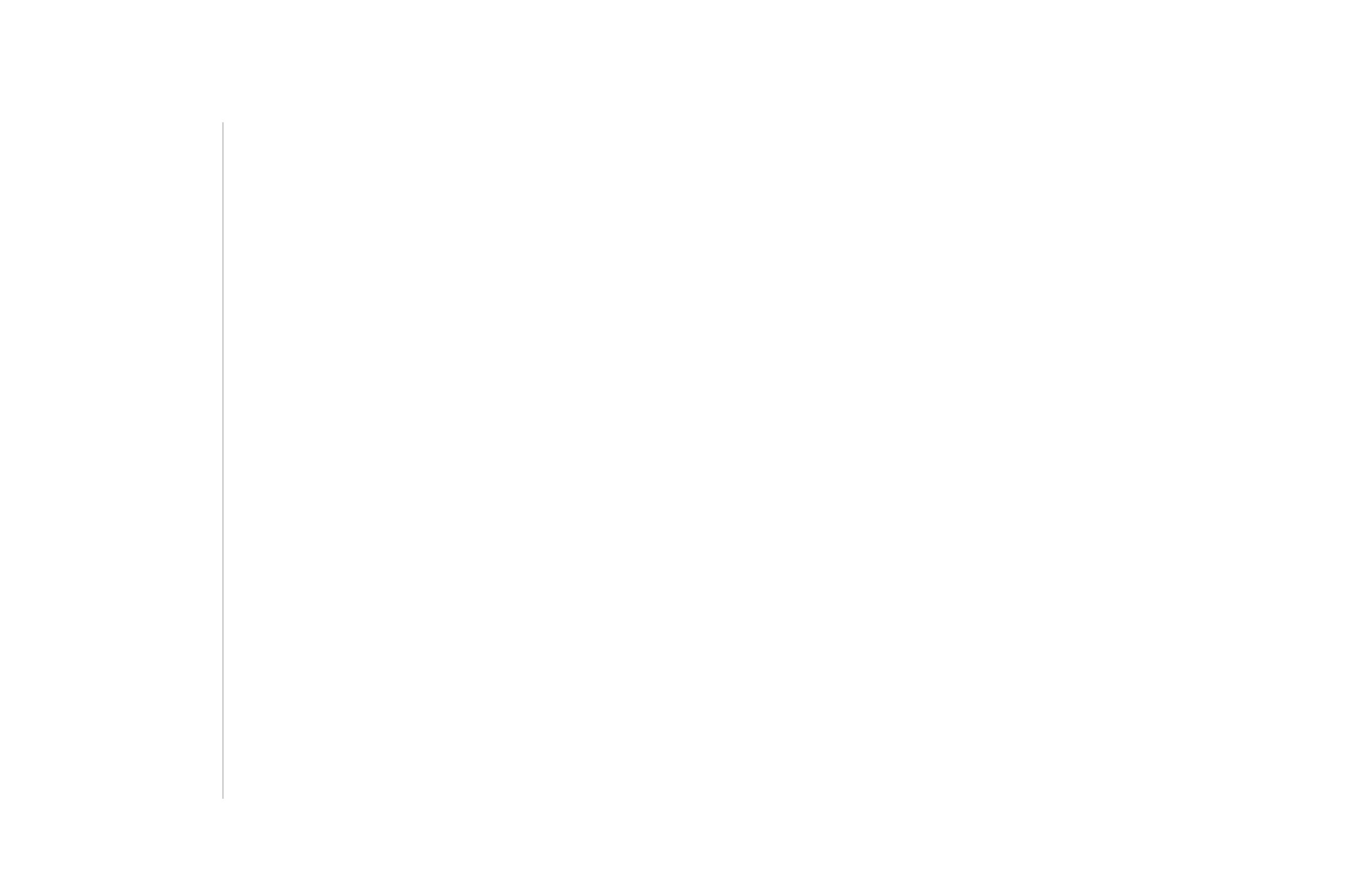
		\end{scriptsize}
		\vspace{-1em}
		\caption{\emph{multi}}%
		\label{fig:multiablation}
	\end{subfigure}
	\hfill
	\begin{subfigure}[b]{0.26\linewidth}  %
		\graphicspath{{figures/}}
		\def\svgwidth{\linewidth}
		\begin{scriptsize}
			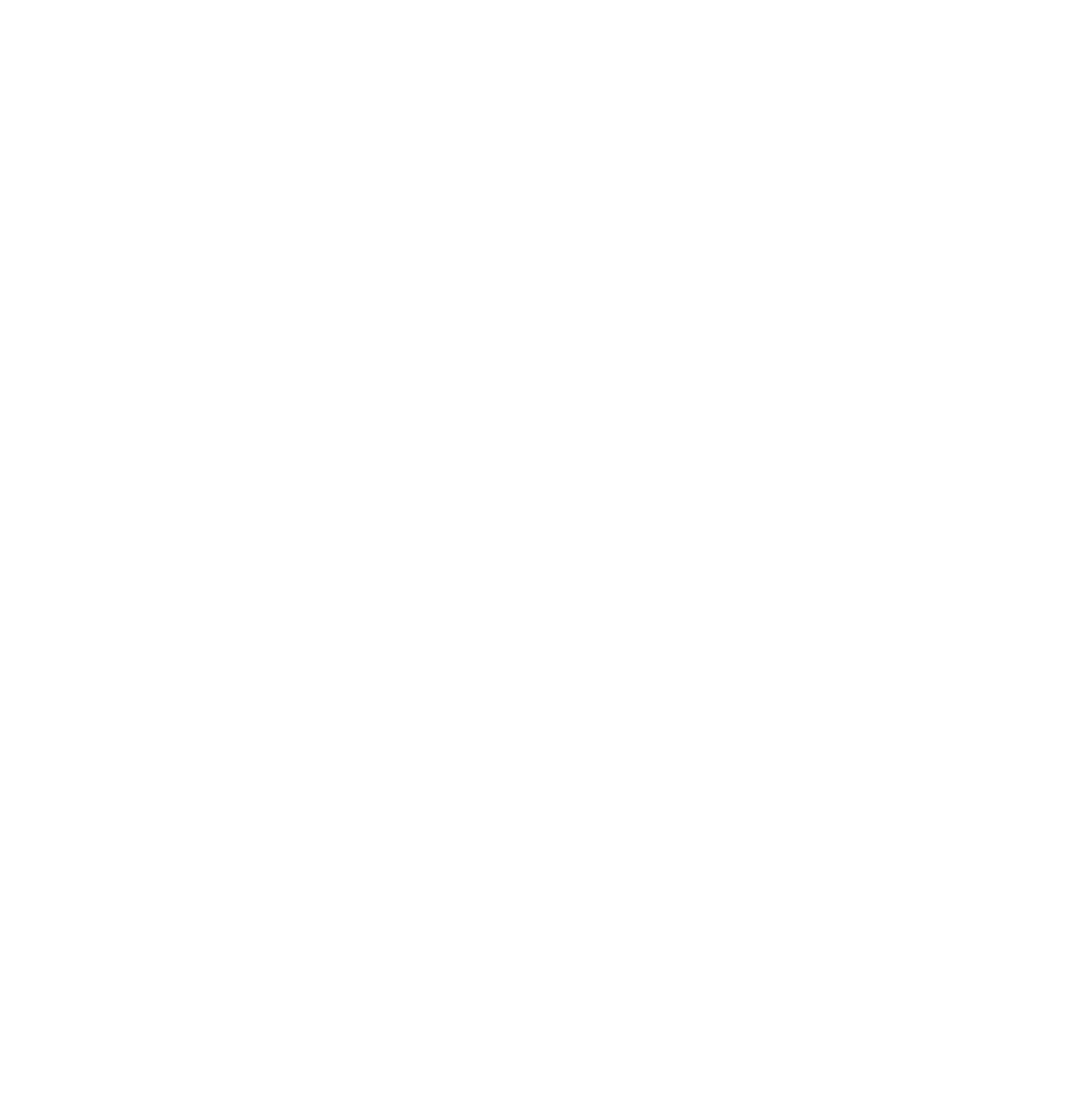
		\end{scriptsize}
		\vspace{-1em}
		\caption{\emph{complex}}
		\label{fig:complexablation}
	\end{subfigure}
	\hfill
	\begin{subfigure}[b]{0.26\linewidth}  %
		\graphicspath{{figures/}}
		\def\svgwidth{\linewidth}
		\begin{scriptsize}
			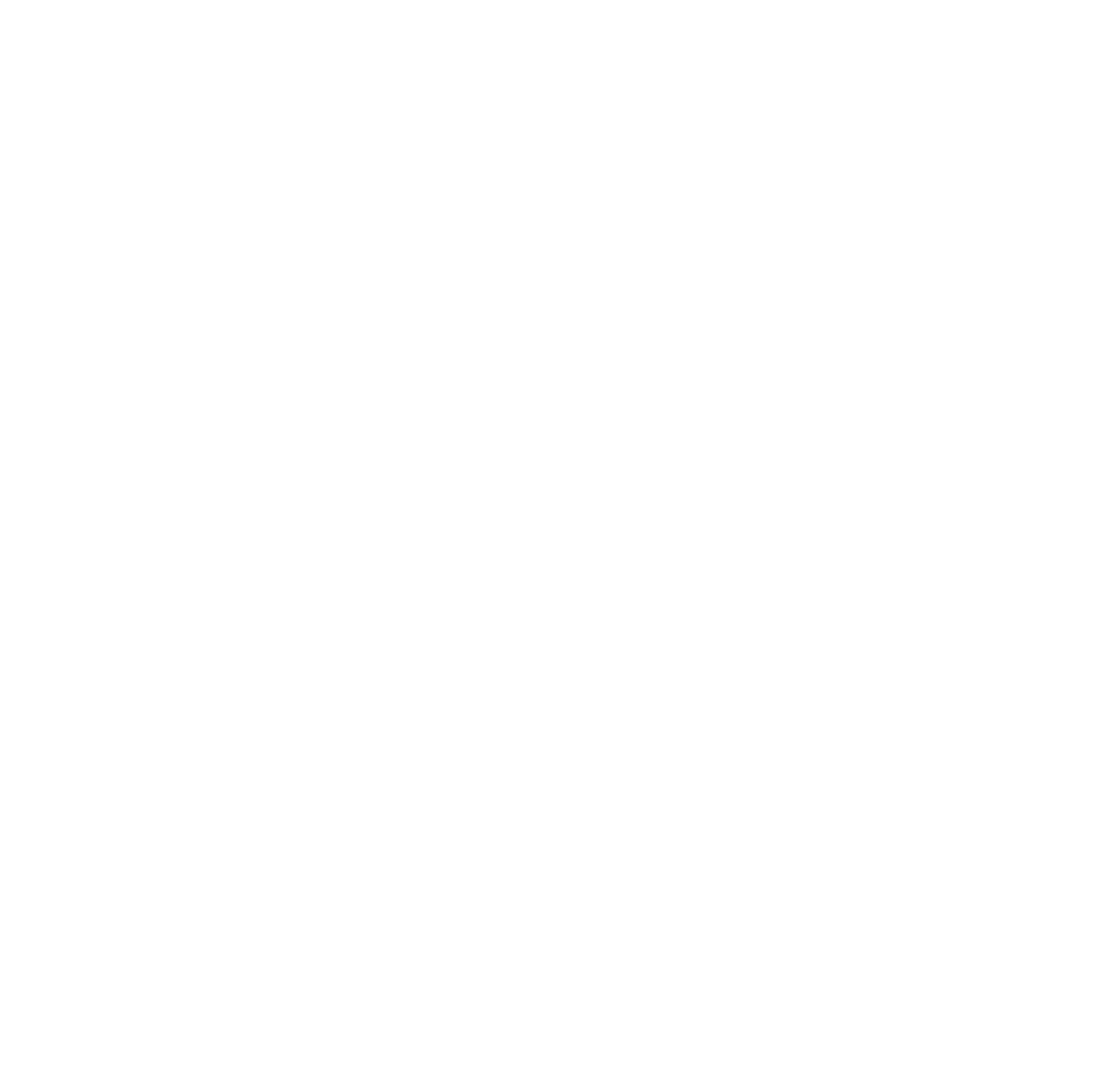
		\end{scriptsize}
		\vspace{-1em}
		\caption{\emph{justlen}}
		\label{fig:justlenablation}
	\end{subfigure}
	\caption{Results of scalability experiments showing the analysis time required over the length of transaction sequences with various configurations of \toolname.}%
	\label{fig:ablationbenches}
\end{figure*}

As part of our ablation study, we also compare the various configurations in terms of scalability to longer and complex transaction sequences.
We utilize the same scalability experiment as in \Cref{sec:eval:scalability}.
However, we bound the execution time to a maximum of \SI{8}{\hour}.
We run \toolname in four configurations: Full \toolname with ABI, full \toolname without knowledge of the ABI, fuzzing with the custom mutator only (\emph{EM}), and fuzzing with AFL++'s mutations only (\emph{AFL}).
The results are shown for the \emph{multi}, \emph{complex}, and \emph{justlen} benchmarks in \Cref{fig:multiablation}, \Cref{fig:complexablation}, and \Cref{fig:justlenablation}, respectively.
While the \emph{AFL} configuration generally provides the highest throughput (see \Cref{tab:throughput}), it lacks structured mutation operations, such as splicing at the transaction level.
As a result, the fuzzing process becomes ineffective and fails to reliably generate even short meaningful transaction sequences.
Furthermore, the \emph{AFL} configuration fails to identify a bug in any of the \emph{complex} contract variants.
This shows that the custom mutator in \toolname is essential for good fuzzing performance.
Furthermore, we can see that in this benchmark, the \emph{EM} performs best since the magic value comparisons are best solved using the dictionary-based sampling employed by the custom mutator for integer types.
Since the custom mutator utilizes the dictionary probabilistically, we also observe some extreme outliers in these experiments.
Since the custom mutator performs mostly structural mutations on the transaction sequences, it also performs best on the \emph{justlen} experiment.

While the \emph{EM} configuration performs best on the benchmarks presented here, we found that without AFL's mutations (especially the input-to-state correspondence~\cite{Aschermann2019redqueen} mutations) there are several fuzzing roadblocks in practice that cannot be solved by the \emph{EM} configuration.
Furthermore, the benchmarks focusing on real-world smart contracts in \Cref{sec:eval:scalability} show that the standard \toolname configuration performs best.

\paragraph{Fuzzing without ABI}
When fuzzing without ABI information, \toolname fully relies on the coverage feedback to discover useful transaction inputs.
Generally, \toolname without ABI information expectedly performs worse compared to fuzzing with ABI information.
On the \emph{multi} and \emph{complex} contracts, fuzzing without ABI information identifies the bug in a mean of \SI{223}{\minute} ($\sigma = 205$).
In comparison, fuzzing with ABI information requires a mean of \SI{87}{\minute} ($\sigma = 127$) to identify the bug.
On the \emph{justlen} benchmark, the difference is much smaller: \SI{3.9}{\minute} with ABI and \SI{6.6}{\minute} without ABI.
To identify a bug in the \emph{justlen} benchmark, the fuzzer does not need to identify correct input parameters, as most exposed functions simply do not require parameters.
Performing structural mutations on the transaction sequence does not require ABI information. 
For this reason, the performance difference is much smaller.

\begin{figure}[t]
	\centering
	\graphicspath{{figures/}}
	\def\svgwidth{0.9\linewidth}
	\begin{scriptsize}
		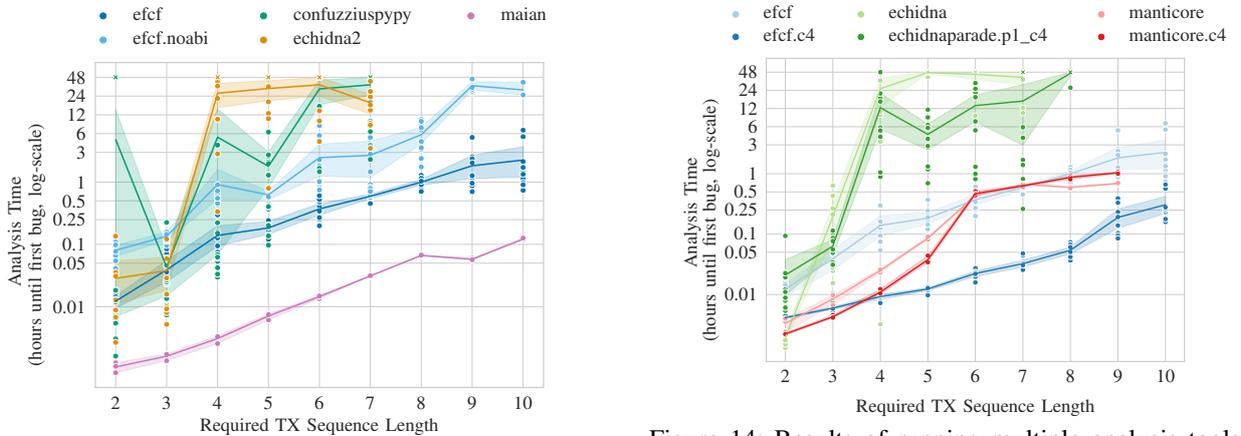
	\end{scriptsize}
	\caption{Comparison of \toolname with and without ABI and other analysis tools. MAIAN is the best analysis tool in this benchmark and never uses ABI information. Echidna and Confuzzius always utilize ABI information. We can see that EF/CF without ABI performs significantly worse but still outperforms the other fuzzers that utilize ABI information.}%
	\label{fig:ablationnoabi}
\end{figure}

The largest difference can be observed in the \emph{multi} benchmark.
This contract primarily tests the analysis tool's capability of finding solutions to multiple input constraints while creating long transaction sequences.
This benchmark heavily favors symbolic execution tools, as there is nearly no potential for state explosion within functions.
\Cref{fig:ablationnoabi} shows a comparison of \toolname with and without ABI with the best performing symbolic execution tool and the fuzzers we evaluated.
While \toolname with ABI features comparable performance to symbolic execution tools, \toolname without ABI performs significantly worse.
Because of the lack of knowledge about the input structure, \toolname samples the dictionary less often than with ABI.
This leads to a decreased chance of placing the correct values into the input.
However, remarkably, \toolname without ABI information still performs significantly better than state-of-the-art fuzzing tools that utilize the ABI.

\section{Additional Benchmarks and Evaluation Details}%
\label{sec:evalappendix}

\paragraph{Multi-Core Performance}
We also evaluate multi-core performance of those analysis tools that support it: \toolname, Manticore, and Echidna.
To parallelize Echidna, we utilize the \emph{echidna-parade}~\cite{Groce2021echidnaparade} tool to run multiple instances of Echidna in parallel.
In contrast to the normal mode of operation in \emph{echidna-parade}, we always fuzz the full set of functions by not excluding any function from the fuzzing runs.
In our benchmarks, the default mode of operation is detrimental to performance in terms of time-to-bug.
Manticore natively supports multi-threaded analysis to leverage multiple cores.
For \toolname, we leverage the multi-core fuzzing approach of AFL++.

\begin{figure}[t]
  \centering
	\graphicspath{{figures/}}
	\def\svgwidth{0.9\linewidth}
	\begin{scriptsize}
		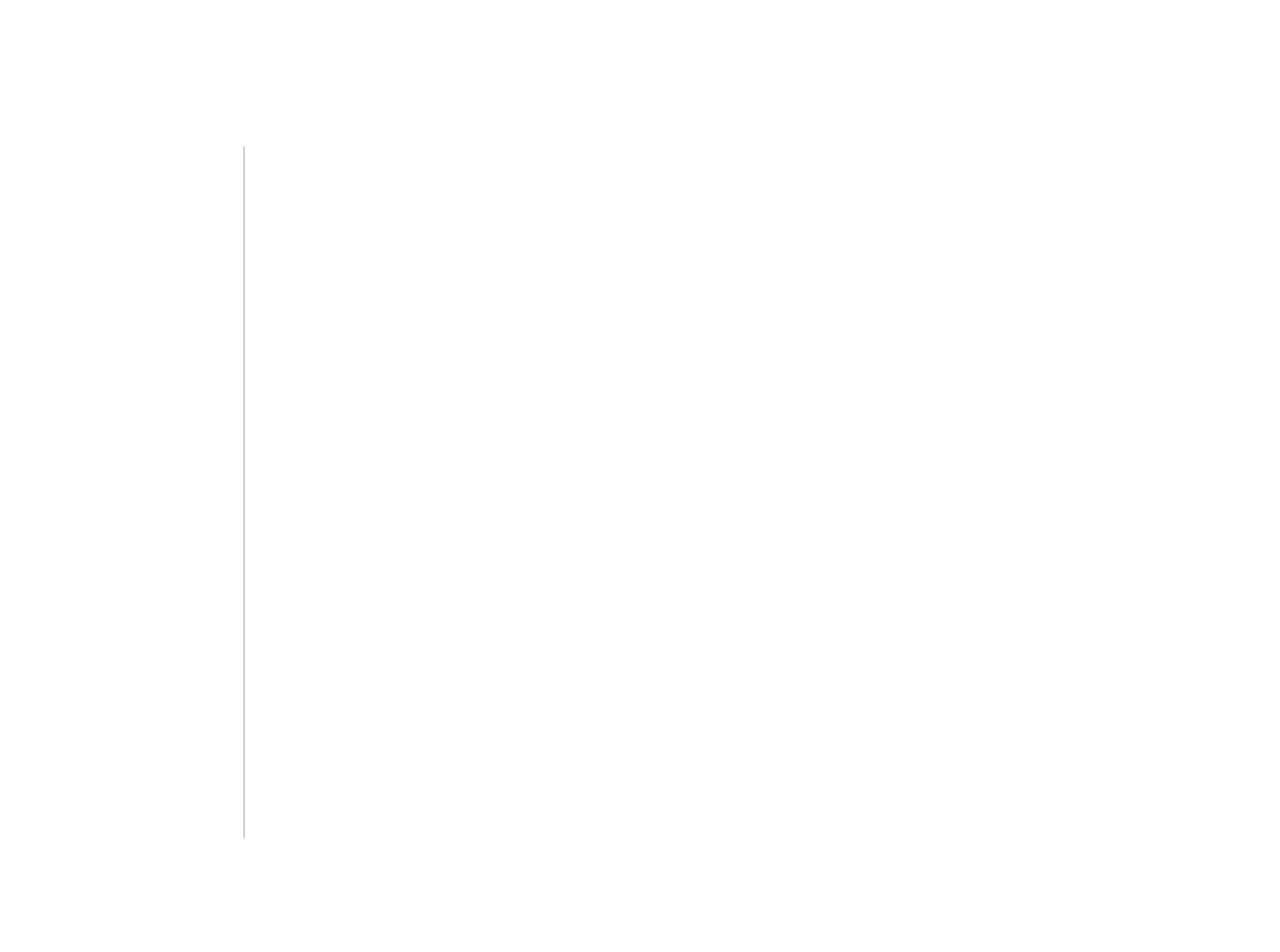
	\end{scriptsize}
  \caption{Results of running multiple analysis tools on a single core vs. running on 4 cores (marked with \emph{c4}) in parallel on the \emph{multi} dataset.}%
  \label{fig:multigenplotmulticore}
\end{figure}

\Cref{fig:multigenplotmulticore} shows the multi-core performance of several analysis tools on the \emph{multi} contracts.
We can see that with \toolname, the performance significantly increases between the single and multi-core versions.
This is primarily because \toolname utilizes an ensemble fuzzing-like approach that spawns \toolname's core fuzzer in multiple different configurations.
Similarly, parallelizing Echidna with the \emph{echidna-parade} tool shows significant improvements over the single-core Echidna.
In a multi-core setting, Echidna can find the transaction sequences with up to 8 transactions and features a significant speed-up for the transaction sequences with lengths 4 to 7.
For both fuzzers, we observe that some single-core runs are as fast or faster than other multi-core runs.
Overall, however, the multi-core runs reduce the variance between the runs, allowing the fuzzer to identify the bug in a given time span more consistently.
Interestingly, Manticore, the only other tool with built-in multi-core support, does not gain a significant speedup in this experimental setup.
We suspect that the symbolic execution approach taken by manticore cannot fully leverage multiple cores.

\paragraph{Code Coverage}
In \Cref{sec:eval:codecov}, we describe an experiment to assess the ability of current fuzzers to reach code coverage on a set of real-world smart contracts.
We describe the dataset in more detail in \Cref{sec:benchdetails}.
The overall results are shown in \Cref{tab:coverage}.
Here, we show the number of targets where one fuzzer outperforms another fuzzer with statistical significance.

\begin{table}[b]
	\center
	\caption{Comparison of all fuzzers on the test set: the number of times fuzzer A outperformed fuzzer B.}%
	\label{tab:coverage}
	\begin{tabular}{lccc}
		\toprule
		\diagbox[width=9em]{Fuzzer A}{Fuzzer B} & \toolname & Confuzzius & ILF \\
		\midrule
		\toolname                               & -         & 141                           & 120 \\
		ConFuzzius                              & 83        & -                             & 105 \\
		ILF                                     & 112       & 136                           & -   \\
		\bottomrule
	\end{tabular}
\end{table}

\section{Fuzzing Reentrancy Honeypots}%
\label{sec:honeypotre}

\citewauthor{Torres2019honeybadger} discussed the phenomenon of honeypot contracts.
These contracts are deployed with source code often available on \href{https://etherscan.io}{etherscan.io} that appears to be vulnerable to, e.g., reentrancy attacks.
However, these contracts are, in fact, a form of scam.
They target malicious actors that search for easily exploitable contracts on the blockchain.
They require the attacker to first invest a number of Ether to later exploit the seemingly vulnerable contract.
However, the code hides a mechanism that prevents exploitation, locking the previously invested Ether of the attacker.
Most of the known reentrancy honeypot contracts use a call to an external library-like contract to revert attack transactions.
The deception works by suggesting that the source code on etherscan also provides the source code for the external contract when, in fact, a different contract is used.

Many of the known honeypot contracts feature very obvious reentrancy vulnerabilities, as these contracts are designed to be easily analyzable (i.e., to lure more people into attempting to attack the honeypot).
Many of those reentrancy honeypot contracts ended up in various datasets of prior studies~\cite{bose2021sailfish,Durieux2020smartbugs}.
In the curated version of the smartbugs dataset, the majority of the contracts identified as vulnerable to reentrancy are, in fact, honeypot contracts.
This dataset contains 19 reentrancy honeypots and 12 other contracts vulnerable to reentrancy.

Honeypot contracts introduce significant bias into datasets.
For example, a tool that detects all honeypot contracts in the curated smartbugs dataset already seems to detect the majority of reentrancy bugs.
However, in reality, all these reentrancy bugs follow the exact same code pattern.
For this reason, we chose to summarize all these cases as \emph{trivial reentrancy} in \Cref{sec:eval:bugdetection}.

The reentrancy honeypots can be analyzed in two ways: by relying on the source code only and by importing code and state data directly from the blockchain.
It is important to distinguish both cases since, in the former, the contract is exploitable, while in the latter, it is not.
We verified that \toolname correctly identifies the reentrancy attacks in the first case (see \Cref{sec:eval:bugdetection}).
Here we deploy a fresh instance of the contract, and the mechanism to prevent exploitation does not work.
In this case, \toolname can correctly identify the reentrancy vulnerability.
However, if we export the contract's state from the blockchain, including the external contract that is called, then the mechanism to prevent exploitation is working.
In this case, \toolname also executes the second external contract, reverting the transaction before the reentrancy takes place.
Thus, \toolname correctly does not report any false alarm.

\section{Sailfish Reentrancy Findings}%
\label{sec:sailfishre}

As part of the evaluation of the \textsc{Sailfish} tool, \citeauthor*{bose2021sailfish} released a list of contracts where reentrancy causes inconsistent state according to \textsc{Sailfish}.
This list contains \num{1904} contracts, of which the \citeauthor*{bose2021sailfish} verified \num{26} to be true positives\footnote{\url{https://github.com/ucsb-seclab/sailfish}}.
Among the list of \num{1904} contracts, \toolname identifies vulnerabilities in only \num{67} contracts.
However, in \num{8} of these \num{67} contracts, \toolname discovers a vulnerability unrelated to reentrancy, e.g., a controlled \emph{delegatecall} vulnerability.

Furthermore, we analyzed the list of verified true positives in more detail.
Among the vulnerable contracts reported by the \textsc{Sailfish} tool, \toolname correctly identifies 5 contracts that can be exploited with reentrancy to steal Ether.
Among these five contracts, one is a test contract, one a known honeypot, and the remaining 3 contracts are duplicates.
Furthermore, \toolname identifies one contract that can be exploited due to an access control bug, not a reentrancy.
Note that \toolname only identifies the honeypot as vulnerable when deploying from source code (see \Cref{sec:honeypotre}).
We manually identified one contract that seems to be vulnerable to reentrancy, but no Ether is at stake.
The remaining contracts exhibit reentrancy patterns but are probably not exploitable.

For example, the \emph{CommonWallet} contract, depicted in \Cref{fig:commonwallet}, is affected by a similar token-related reentrancy as the \emph{Uniswap-V2} contract~\cite{Torres21eyeofhorus}.
Here, the attacker must supply an \emph{ERC777} token where an \emph{ERC20} token is expected.
Most (legitimate) \emph{ERC20} token contracts do not perform callbacks to the attacker and therefore the attacker cannot trigger a reentrancy situation.
However, this is different for \emph{ERC777} contracts that feature callbacks by design.
They allow the attacker to reenter the \emph{CommonWallet} contract and trigger a reentrancy.
Currently, \toolname does not detect token-related bugs since \toolname has no concept of tokens and therefore does not regard token gains as a bug.

However, in reality, this reentrancy bug cannot be used to cause damage.
The reason for this is that the attacker would have to cause an integer underflow, which is prevented due to the integer checking leveraged by the contract.
\toolname would not identify a possible reentrancy attack, even if there were a bug oracle for tokens, because there is no way to exploit the reentrancy attack.
We verified this by testing \toolname with a contract that exhibits the same vulnerability but with Ether instead of tokens.
We believe the reason for this false alarm is that Sailfish does not accurately model the transaction-based execution model of the EVM.
It will eagerly report a bug without considering that the transaction will revert later, a flaw common to many analysis tools~\cite{Schneidewind2020goodbadugly}.

\begin{figure}[th]
\begin{minted}[%
    frame=lines,
    framesep=2mm,
    %
    fontsize=\footnotesize,
    stripnl=false,
    linenos,
    xleftmargin=1.8em,
		breaklines=true,
    ]{solidity}
function safeSub(uint256 _x, uint256 _y) internal pure returns (uint256) {
  assert(_x >= _y);
  return _x - _y;
}
function sendTokenTo(address tokenAddr, address to_, uint256 amount) {
  require(tokenBalance[tokenAddr][msg.sender] >= amount);
  /* external call - might cause reentrancy */
  if(ERC20Token(tokenAddr).transfer(to_, amount)) {
    /* state update with underflow check in safeSub */
    tokenBalance[tokenAddr][msg.sender] =
        safeSub(tokenBalance[tokenAddr][msg.sender], amount);
  }
}
\end{minted}
  \caption{\emph{CommonWallet} reentrancy, which is not exploitable due to the integer overflow check.}%
  \label{fig:commonwallet}
\end{figure}

Interestingly, Sailfish uses a rather broad definition of state inconsistency caused by reentrancy.
Any reachable state write that could cause inconsistency is considered a true alarm.
We found several contracts where reentrancy is possible but where the reentrancy will never actually cause inconsistent state.
Here, a state variable is written but never changed, e.g., the reentrant code path will perform a subtraction with 0. 
However, this state variable update is identified by \textsc{Sailfish} to cause state inconsistency.
In contrast, \toolname will only report bugs that actually cause damage.
Furthermore, as a static analysis tool, Sailfish must over-approximate and consider all external calls as being able to cause reentrancy.
However, in practice, many contracts issue calls to trusted contracts that do not allow the attacker to perform reentrancy.
In contrast, \toolname will not perform reentrancy on calls to unknown contracts by default, avoiding false reentrancy alarms.
For highest precision, it can also import and execute trusted contract dependencies of the target.

\begin{figure}[t]
  \centering
	\inputminted{yaml}{figures/ReentrancyVulnBankBuggyLockHard.attack.yml}
  \caption{Textual representation of the transaction sequence generated by \toolname to exploit the contract from \Cref{fig:hardre}.}%
  \label{fig:hardeattack}
\end{figure}

\end{document}